\documentclass[10pt, a4paper]{article}

\usepackage{geometry}
\usepackage{mathtools}
\usepackage{amsmath}
\usepackage{amssymb}
\usepackage{amsfonts}
\usepackage{amsthm}
\usepackage{relsize}

\usepackage[english]{babel}
\usepackage[utf8]{inputenc}
\usepackage[T1]{fontenc}
\usepackage[hidelinks]{hyperref}
\usepackage[round]{natbib}
\usepackage[pdftex,dvipsnames,table]{xcolor}
\usepackage[nameinlink,noabbrev]{cleveref}

\usepackage{float}
\usepackage{environ}
\usepackage{tikz} 
\usepackage{tikzsymbols}

\usepackage{indentfirst}
\usepackage{url}
\usepackage{latexsym}
\usepackage{calc}
\usepackage{epsfig}
\usepackage{fancyhdr}
\usepackage{graphicx}
\usepackage{scalerel}
\usepackage{xargs}
\usepackage{xspace}
\usepackage{pdftexcmds}
\usepackage{pgfplots}
\usepackage{hyphenat}

\usepackage{etoolbox}
\usepackage{setspace}
\usepackage{bookmark}
\usepackage{booktabs}
\usepackage{multirow}

\usepackage[usestackEOL]{stackengine}

\edef\restoreparindent{\parindent=\the\parindent\relax}
\usepackage{parskip}
\restoreparindent
\usepackage[font={small},labelfont=bf]{caption}
\usepackage{subcaption}

\usepackage[shortlabels]{enumitem}

\usepackage[noend]{algpseudocode}
\usepackage{algorithm}

\usepackage{nicefrac}

\algrenewcommand{\algorithmicwhile}{\textbf{While}}
\algrenewcommand{\algorithmicif}{\textbf{If}}
\algrenewcommand{\algorithmicelse}{\textbf{Else}}

\algnewcommand{\algorithmicforeach}{\textbf{For each}}
\algdef{S}[FOR]{ForEach}[1]{\algorithmicforeach\ #1\ \algorithmicdo}

\usepackage{soul}

\usepackage{accents}
\newcommand{\ubar}[1]{\underaccent{\bar}{#1}}

\usepackage{lmodern}
\usepackage{microtype}
\DisableLigatures{encoding = *, family = *}

\newcommand\Times{\scaleobj{0.6}{\times}}

\usepackage{etoolbox}
\usepackage{lineno}
\usepackage[normalem]{ulem}
\usepackage[colorinlistoftodos,prependcaption,textsize=tiny]{todonotes}

\newcommandx{\unsure}[2][1=]{%
  \todo[linecolor=red, backgroundcolor=red!25, bordercolor=red, #1]{%
  \begin{flushleft}\setstretch{1.25}#2\end{flushleft}%
}}

\newcommandx{\change}[2][1=]{%
  \todo[linecolor=blue, backgroundcolor=blue!25, bordercolor=blue, #1]{%
  \begin{flushleft}\setstretch{1.25}#2\end{flushleft}%
}}

\newcommandx{\info}[2][1=]{%
  \todo[linecolor=OliveGreen, backgroundcolor=OliveGreen!25, bordercolor=OliveGreen, #1]{%
  \begin{flushleft}\setstretch{1.25}#2\end{flushleft}%
}}

\newcommandx{\improvement}[2][1=]{%
  \todo[linecolor=Plum, backgroundcolor=Plum!25, bordercolor=Plum, #1]{%
  \begin{flushleft}\setstretch{1.25}#2\end{flushleft}%
}}

\newcommandx{\thiswillnotshow}[2][1=]{\todo[disable, #1]{#2}}

\definecolor{RED}{rgb}{1,0,0}\definecolor{BLUE}{rgb}{0,0,1}

\providecommand{\DIFdel}[1]{} 

\makeatletter
\newif\iftx@libertine
\newif\iftx@minion
\newif\iftx@coch
\newif\iftx@ch
\newif\iftx@stxtwo
\newif\iftx@ebgm
\newif\iftx@ut
\newif\iftx@nc
\newif\iftx@ams

\newcommand{\appendixnumberline}[1]{Appendix\space}

\def\mylabel#1{\ifmeasuring@\else\ltx@label{#1}\fi}
\DeclareRobustCommand{\makeref}[1]{%
  \makephantom%
  \quad\label{#1}
}

\DeclareRobustCommand{\veb}[1]{\mathpalette\do@veb{#1}}
\newcommand{\do@veb}[2]{%
  \fix@cev{#1}{+}%
  \mbox{$\m@th#1\vec{\mbox{$\fix@cev{#1}{-}\m@th#1#2\fix@cev{#1}{+}$}}$}%
  \fix@cev{#1}{-}%
}

\DeclareRobustCommand{\cev}[1]{\mathpalette\do@cev{#1}}
\newcommand{\do@cev}[2]{%
  \fix@cev{#1}{+}%
  \reflectbox{$\m@th#1\vec{\reflectbox{$\fix@cev{#1}{-}\m@th#1#2\fix@cev{#1}{+}$}}$}%
  \fix@cev{#1}{-}%
}

\newcommand{\fix@cev}[2]{%
  \ifx#1\displaystyle 
    \mkern#23mu
  \else
    \ifx#1\textstyle
      \mkern#23mu
    \else
      \ifx#1\scriptstyle
        \mkern#22mu
      \else
        \mkern#22mu
      \fi
    \fi
  \fi
}

\algrenewcommand\ALG@beginalgorithmic{\small} 

\AfterEndEnvironment{algorithm}{\let\@algcomment\relax}
\AtEndEnvironment{algorithm}{\kern2pt\hrule\relax\vskip3pt\@algcomment}
\let\@algcomment\relax
\newcommand\algcomment[1]{\def\@algcomment{\footnotesize#1}}

\renewcommand\fs@ruled{\def\@fs@cfont{\bfseries}\let\@fs@capt\floatc@ruled
  \def\@fs@pre{\hrule height.8pt depth0pt \kern2pt}%
  \def\@fs@post{}%
  \def\@fs@mid{\kern2pt\hrule\kern2pt}%
  \let\@fs@iftopcapt\iftrue%
}

\newcommand{\leqnomode}{\tagsleft@true\let\veqno\@@leqno}
\newcommand{\reqnomode}{\tagsleft@false\let\veqno\@@eqno}

\def\th@plain{%
  \thm@notefont{}
  \itshape 
}
\def\th@definition{%
  \thm@notefont{}
  \normalfont 
}

\makeatother

\algnewcommand{\NewComment}[1]{%
  \textcolor{OliveGreen}{%
  \fontfamily{ppl}\selectfont%
    \(\triangleright\)~\footnotesize{#1}%
  }%
}

\algnewcommand{\LeftComment}[1]{%
  \Statex\NewComment{#1}%
}

\newcommand{\RemoveSpaces}[1]{%
  \begingroup%
  \spaceskip=1.2mm%
  \xspaceskip=1.2mm%
  \mbox{#1}%
  \endgroup
}

\DeclareMathAlphabet{\matheuler}{U}{zeur}{m}{n}
\DeclareSymbolFont{lettersslanted}{OML}{zplm}{m}{n}
\DeclareMathSymbol{p}{\mathalpha}{lettersslanted}{`p}

\newcommand{\shortname}{EVSP\xspace}
\newcommand{\ie}{i.e.\xspace}
\newcommand{\eg}{e.g.\xspace}

\newcommand{\incoming}[1]{\!\smash{\cev{\,#1}}}
\newcommand{\outgoing}[1]{\!\!\smash{\veb{\,\,#1}}}

\newcommand{\basephantom}{\incoming{\matheuler{E}}}
\newcommand{\makephantom}{\vphantom{\sum\limits_{\mathrlap{\substack{\basephantom}}}}}
\newcommand{\centerlap}[1]{\limits_{\mathclap{\substack{#1\vphantom{\basephantom}}}}}
\newcommand{\rightlap}[1]{\limits_{\mathrlap{\substack{\mathsurround=-7pt#1\vphantom{\basephantom}}}}}

\newcommand{\lowprime}{\mkern0mu\raise0.6ex\hbox{$\scriptstyle\prime$}}
\newcommand{\CC}{%
  \sffamily C\nolinebreak\hspace{-.01em}\raisebox{.3ex}{\footnotesize +}%
             \nolinebreak\hspace{-.01em}\raisebox{.3ex}{\footnotesize +}%
}

\newcommand{\eqrefsmallsize}[1]{\eqref{#1}}

\newcounter{fc}
\newcounter{sc}

\renewcommand{\thefc}{\shortname\!\!\arabic{fc}}
\renewcommand{\thesc}{\shortname\!\!\arabic{sc}-S}

\newcommand{\makerefF}[1]{\vspace*{0.15cm}\refstepcounter{fc}%
  \noindent(\hspace*{.2mm}\thefc\label{#1})\xspace
}

\newcommand{\makerefSb}[1]{\refstepcounter{sc}%
  \thesc\label{#1}\xspace
}

\newtheorem{theorem}{Theorem}

\newtheorem{corollary}{Corollary}
\newtheorem{proposition}{Proposition}

\newtheorem{remark}{Remark}

\newcommand{\textBF}[1]{\pdfliteral direct {2 Tr 0.3 w}#1\pdfliteral direct {0 Tr 0 w}}

\pgfplotsset{
  box plot/.style={
    /pgfplots/.cd,
    black,
    only marks,
    mark=-,
    mark size=\pgfkeysvalueof{/pgfplots/box plot width},
    /pgfplots/error bars/y dir=plus,
    /pgfplots/error bars/y explicit,
    /pgfplots/table/x index=\pgfkeysvalueof{/pgfplots/box plot x index},
  },
  box plot box/.style={
    /pgfplots/error bars/draw error bar/.code 2 args={%
      \draw  ##1 -- ++(\pgfkeysvalueof{/pgfplots/box plot width},0pt) %
         |- ##2 -- ++(-\pgfkeysvalueof{/pgfplots/box plot width},0pt) |- ##1 -- cycle;
    },
    /pgfplots/table/.cd,
    y index=\pgfkeysvalueof{/pgfplots/box plot box top index},
    y error expr={
      \thisrowno{\pgfkeysvalueof{/pgfplots/box plot box bottom index}}
      - \thisrowno{\pgfkeysvalueof{/pgfplots/box plot box top index}}
    },
    /pgfplots/box plot
  },
  box plot top whisker/.style={
    /pgfplots/error bars/draw error bar/.code 2 args={%
      \pgfkeysgetvalue{/pgfplots/error bars/error mark}%
      {\pgfplotserrorbarsmark}%
      \pgfkeysgetvalue{/pgfplots/error bars/error mark options}%
      {\pgfplotserrorbarsmarkopts}%
      \path ##1 -- ##2;
    },
    /pgfplots/table/.cd,
    y index=\pgfkeysvalueof{/pgfplots/box plot whisker top index},
    y error expr={
      \thisrowno{\pgfkeysvalueof{/pgfplots/box plot box top index}}
      - \thisrowno{\pgfkeysvalueof{/pgfplots/box plot whisker top index}}
    },
    /pgfplots/box plot
  },
  box plot bottom whisker/.style={
    /pgfplots/error bars/draw error bar/.code 2 args={%
      \pgfkeysgetvalue{/pgfplots/error bars/error mark}%
      {\pgfplotserrorbarsmark}%
      \pgfkeysgetvalue{/pgfplots/error bars/error mark options}%
      {\pgfplotserrorbarsmarkopts}%
      \path ##1 -- ##2;
    },
    /pgfplots/table/.cd,
    y index=\pgfkeysvalueof{/pgfplots/box plot whisker bottom index},
    y error expr={
      \thisrowno{\pgfkeysvalueof{/pgfplots/box plot box bottom index}}
      - \thisrowno{\pgfkeysvalueof{/pgfplots/box plot whisker bottom index}}
    },
    /pgfplots/box plot
  },
  box plot median/.style={
    /pgfplots/box plot,
    /pgfplots/table/y index=\pgfkeysvalueof{/pgfplots/box plot median index}
  },
  box plot width/.initial=1em,
  box plot x index/.initial=0,
  box plot median index/.initial=1,
  box plot box top index/.initial=2,
  box plot box bottom index/.initial=3,
  box plot whisker top index/.initial=4,
  box plot whisker bottom index/.initial=5,
  x tick label style={/pgf/number format/.cd,fixed,precision=3, set thousands separator={}}
}

\usetikzlibrary{spy,positioning,arrows,arrows.meta}
  
\newcommand{\boxplot}[2][]{
  \addplot [box plot median,#1] table {#2};
  \addplot [forget plot, box plot box,#1] table {#2};
  \addplot [forget plot, box plot top whisker,#1] table {#2};
  \addplot [forget plot, box plot bottom whisker,#1] table {#2};
}

\title{\LARGE%
On the approximability and energy-flow modeling of the electric vehicle sharing problem
}

\author{Welverton R. Silva, Fábio L. Usberti, Rafael C.\,S. Schouery\thanks{Corresponding author.%
}}

\begin{document}
\date{}
\maketitle

\vspace{-20pt}
\begin{center}\footnotesize 
  Institute of Computing, University of Campinas\\Campinas, São Paulo, Brazil
\end{center}


\bigskip\bigskip\noindent
{\small {\bf ABSTRACT.}
The electric vehicle sharing problem~(\shortname) arises from the planning and operation of one-way electric car-sharing systems.~It aims to maximize the total rental time of a fleet of electric vehicles while ensuring that all of the demands of the customer are fulfilled.~In this paper, we expand the knowledge on the complexity of the EVSP by showing that it is~\mbox{NP-hard} to approximate it to within a factor of~\(n^{1-\epsilon}\)\! in polynomial time, for any~\mbox{\(\epsilon > 0\)}, where~\(n\) denotes the number of customers, unless~\mbox{P = NP}.~In addition, we also show that the problem does not have a monotone structure, which can be detrimental to the development of heuristics employing constructive strategies.~Moreover, we propose a novel approach for the modeling of the~\shortname based on energy flows in the network.~Based on the new model, we propose a relax-and-fix strategy and an exact algorithm that uses a warm-start solution obtained from our heuristic approach.~We report computational results comparing our formulation with the best-performing formulation in the literature.~The results show that our formulation outperforms the previous one concerning the number of optimal solutions obtained, optimality gaps, and computational times.~Previously,~\(32.7\%\) of the instances remained unsolved (within a time limit of one hour) by the best-performing formulation in the literature, while our formulation obtained optimal solutions for all instances.~To stress our approaches, two more challenging new sets of instances were generated, for which we were able to solve~\(49.5\%\) of the instances, with an average optimality gap of~\(2.91\%\) for those not solved optimally.


\baselineskip=\normalbaselineskip
\sloppy

\section{Introduction}
\label{sec:introduction}

Shared mobility is an emerging segment of transportation systems that advocates for the use of vehicles shared by many users, either concurrently or sequentially.~Car-sharing is one type of shared mobility that rents cars for short periods, low costs, for multiple users, in a single day.~It can improve the quality of urban transport systems by reducing congestion and the need for additional parking spaces~\citep{Shaheen/2007,Zakaria/2014}.~As an added social benefit, car-sharing services also minimize the urban concentration of polluting particles in the air if the cars are powered by electricity.~In this sense, electric car-sharing represents a transition to more sustainable models of urban transportation~\citep{Gambella/2018}.

In station-based car-sharing systems, users can only pick up and drop off rental cars at dedicated stations.~Such systems provide two main types of sharing models, namely \emph{two-way} and \emph{one-way}.~Users must pick up and drop off the car at the same station if the system requires two-way trips.~If the system allows one-way trips, however, users can return the car to any station of their choice as long as the drop-off station and time are specified in advance~\citep{Nourinejad/2015:May}.

In this work, we consider a station-based electric car-sharing system that allows one-way trips.~On this matter, we focus on a combinatorial optimization problem that arises from the planning of driving operations called the electric vehicle sharing problem~(\shortname), proposed by~\cite{Silva/2023}, and described next.~Given a set of stations, a set of~(electric) vehicles distributed at the stations, and a set of customers with one or more driving demands, find an assignment plan~(as a space-time route) for each vehicle such that~(i) whenever one customer's demand is fulfilled, all of their other demands are also fulfilled,~(ii) no vehicle runs out of charge before the customer reaches the drop-off station,~(iii) no station exceeds the number of parked vehicles, and~(iv) the total rental time of the assignment plans for the vehicles is maximized.

The main difference between the~\shortname and other electric car-sharing problems related in the operations research literature is the \emph{customers' satisfaction criteria}, that is, whenever one customer's driving demand is fulfilled, all their other demands must also be fulfilled.

\Cref{fig:example_of_multiple_demands} illustrates an example where a customer has four driving demands.~In this example, commuting to work, there is one demand to drive from station~\textsf{A} (in a residential area) to station~\textsf{B}, located near a subway system, and another demand to drive from station~\textsf{C} to station~\textsf{D} (in a central business district).~After a working day, the customer returns to their home with two additional driving demands.

\begin{figure}[H]
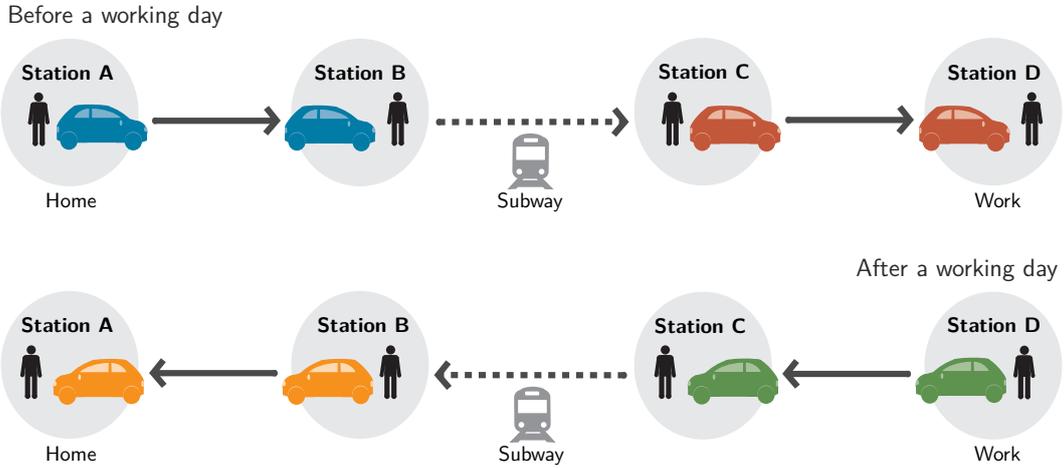

  \centering
  \resizebox{.97\textwidth}{!}{%
    \input{tikz/One-Way_A}
  }
  \resizebox{.97\textwidth}{!}{%
    \input{tikz/One-Way_B}
  }
  \caption{An example of a customer with four driving demands in a one-way car-sharing system.}
  \label{fig:example_of_multiple_demands}
\end{figure}

In that case, not guaranteeing that each of the four customer's driving demands will be fulfilled could be detrimental to their needs.~For the car-sharing service provider, in a general context, customers canceling their driving demands as a result of having some of their demands not guaranteed to be fulfilled can have a substantial negative impact on the planning of trips, which includes routes and timetables of the cars.


\subsection{Previous and related work}

The majority of the studies addressing combinatorial optimization in the literature have focused on vehicle relocation operations~\citep{Ait-Ouahmed/2018, Boyaci/2015, Boyaci/2017, Bruglieri/2014, Bruglieri/2017, Gambella/2018, Huang/2020, Qin/2022, Zhao/2018}; location for changing stations~\citep{Brandstatter/2017, Calik/2019, Deza/2022}; and fleet size planning~\citep{George/2011, Xu/2019}.~In this subsection, we present a brief review of the literature strictly related to our problem.

The~\shortname is a variant of the car-sharing assignment problem first studied by~\cite{Bohmova/2016}, which aims to maximize the number of served customers.~For a complexity analysis, the authors formally define a particular case of the problem with two stations, called~\emph{transfers for commuting}~(TFC), in which each customer has exactly two driving demands that are between the two stations~(in opposite directions).~They prove that the~TFC problem is an~\mbox{NP-hard} problem, and also an~\mbox{APX-hard} problem, even if there is only one car and all of the demands for rent cars take the same amount of time to move between the two stations.

\cite{Silva/2020} presented a mixed-integer linear programming formulation for the~TFC problem, which is modeled as a network flow problem over time.~In their work, the authors propose a preprocessing procedure to reduce the network size, which implies decreasing the number of binary variables and constraints in the mixed-integer linear program.~They also provide an additional set of constraints to strengthen the formulation, which can eliminate integer solutions, but that preserves at least one optimal solution.~To demonstrate the effectiveness of their approaches, computational experiments were conducted on random instances.

\cite{Luo/2020} address the on-line version of the~TFC problem, where customers submit a pair of demands and the scheduler must decide whether to accept each pair of demands immediately at the time when it is submitted.~The authors investigate the impact of different constraints on the booking time of demand pairs and provide lower bounds on the competitive ratio.~Furthermore, they propose a~\(4\)-competitive algorithm for this problem with variable booking times and slotted pick-up times.~For a study of two variants of this on-line version, see the work of~\cite{Luo/2022}.

The~\shortname is introduced by~\cite{Silva/2023}, to optimize the operation of a fleet of electric vehicles for one-way car-sharing systems.~The authors propose four mixed-integer linear programming formulations, based on two main ideas, namely homogeneous and heterogeneous space-time networks.~Also, the proposed formulations are compared theoretically, in terms of the strength of their linear programming relaxation, and computationally, in terms of runtime and solution quality, using benchmark instances based on real data from the~VAMO Fortaleza system~\citep{Fortaleza/2018}.

\subsection{Problem description}

Hereafter, we apply the same notation adopted by~\cite{Silva/2023}, as described next.~In the following, consider the superscript~\(\outgoing{~}\)\, as an ``outgoing'' labeling and the superscript~\(\incoming{~}\)\, as an ``incoming'' labeling when specifying that something (\eg, a vehicle) is leaving or arriving at a location or position~(\eg, a station), respectively.

Let~\(\matheuler{C}\) and~\(\matheuler{S}\) be the sets of customers and stations, respectively, and let~\mbox{\(\matheuler{T} = \{t_{1}, \dots, t_{m}\}\)} be a set of discrete-time instants, not necessarily uniformly distributed, to be considered in a planning horizon~(given in minutes).

For each customer~\mbox{\(c \in \matheuler{C}\)}, let~\(\matheuler{D}_{c}\) be the set of demands for customer~\(c\).~All demand in~\(\matheuler{D}_{c}\) is described by a quintuple~\mbox{\((\outgoing{s}, t_{i}, \incoming{s}, t_{j}, \varepsilon) \in (\matheuler{S} \times \matheuler{T})^2 \times \mathbb{Q}_{>0}\)} such that~\mbox{\(t_{i} < t_{j}\)}, with~\(\outgoing{s}\)~as the pick-up station,~\(t_{i}\) as the departure time,~\(\incoming{s}\) as the drop-off station,~\(t_{j}\) as the arrival time, and~\(\varepsilon\) as the estimate of the energy required during the rental.~In a set~\(\matheuler{D}_{c}\) of demands, there are no overlapping rental periods.~In addition, all demands from customers are grouped into one set.~Let~\mbox{\(\matheuler{D} = \bigcup_{c \in \matheuler{C}} \matheuler{D}_{c}\)} denote the set of all demands.

For each station~\mbox{\(s \in \matheuler{S}\)}, let~\(\matheuler{V}_{s}\) be the set of~EVs initially located at~\(s\).~Each station~\mbox{\(s \in \matheuler{S}\)} has a capacity~\(\textsf{C}_{s}\) and a number~\(\textsf{R}_{s}\) of charging facilities, such that~\mbox{\(\textsf{R}_{s} \leq \textsf{C}_{s}\)} and~\mbox{\(|\matheuler{V}_{s}| \leq \textsf{C}_{s}\)}.~Let~\mbox{\(\matheuler{V} = \bigcup_{s \in \matheuler{S}} \matheuler{V}_{s}\)} denote the set of all~EVs in the system.

All~EVs and charging facilities are identical, in which \mbox{\(\mu \in \mathbb{Q}_{>0}\)} represents the energy supplied by a charging facility per time unit~(in watts per minute), and~\(\textsf{L}\) represents the battery capacity~(in watt-minutes).~Additionally, let~\(\textsf{L}^{v}\) denote the amount of energy that is stored in the battery of vehicle~\(v \in \matheuler{V}\) at the beginning of the planning horizon.

The objective of the~\shortname is to maximize the total rental time by assigning a sequence of driving demands to the fleet of vehicles, associated with a feasible assignment plan, such that whenever one customer's demand is fulfilled, all other customer's demands are also fulfilled.

\begin{remark}
  \normalfont An~\emph{assignment plan} for a vehicle~\mbox{\(v \in \matheuler{V}\)} is feasible if and only if, for each demand~\mbox{\((\outgoing{s}, t_{i}, \incoming{s}, t_{j}, \varepsilon)\)} in a sequence of driving demands to be fulfilled, it satisfies the following conditions: the vehicle \(v\) is at the pick-up station \(\outgoing{s}\) at time~\(t_{i}\); the battery energy of~\(v\) is greater than or equal to the required energy~\(\varepsilon\); and there is a parking space at the drop-off station that is available immediately before time~\(t_{j}\).~Moreover, the assignment plan must specify if and when the vehicle~\(v\) is connected to a charging facility.
\end{remark}

To better clarify the~\shortname, we give an example with three stations, each with a capacity for two vehicles, such that one parking space is equipped with a charging facility while the other is not, and there is only one (fully charged) vehicle initially available.~The vehicle's battery has a capacity of~\(30\) kilowatt-hours~(kWh), and the charging facility provides~\(0.17\)~kWh of energy per minute.~The customers' driving demands are shown in~\Cref{tab:reservation}.

\vspace*{-2mm}
\begin{table}[H]
  \singlespacing
  \caption{Details of the reservation requests for rental.}
  \label{tab:reservation}
  \vspace*{.5mm}
  \resizebox{\textwidth}{!}{%
  \begin{tabular}{ccrcrr}
  \toprule
  Customer & Pick-up station & \multicolumn{1}{c}{Departure time} & Drop-off station & \multicolumn{1}{c}{Arrival time} & \multicolumn{1}{c}{Energy required (kWh)} \\ \hline
  $c_1$    & $s_1$           &  7:28  {~~~~~~~}                    & $s_2$           &  8:06   {~~~~~}                  &  6.34   {~~~~~~~~~~~~~~~~}                \\
  $c_1$    & $s_2$           & 11:23  {~~~~~~~}                    & $s_1$           & 11:54   {~~~~~}                  &  5.17   {~~~~~~~~~~~~~~~~}                \\
  $c_2$    & $s_3$           &  8:35  {~~~~~~~}                    & $s_3$           &  9:32   {~~~~~}                  &  9.00   {~~~~~~~~~~~~~~~~}                \\
  $c_3$    & $s_1$           &  9:46  {~~~~~~~}                    & $s_3$           & 10:12   {~~~~~}                  &  4.33   {~~~~~~~~~~~~~~~~}                \\
  $c_4$    & $s_2$           &  7:40  {~~~~~~~}                    & $s_1$           &  8:02   {~~~~~}                  &  3.67   {~~~~~~~~~~~~~~~~}                \\
  $c_4$    & $s_1$           &  8:54  {~~~~~~~}                    & $s_3$           & 10:28   {~~~~~}                  & 15.66   {~~~~~~~~~~~~~~~~}                \\
  $c_5$    & $s_2$           &  7:28  {~~~~~~~}                    & $s_1$           &  8:23   {~~~~~}                  &  9.17   {~~~~~~~~~~~~~~~~}                \\
  $c_6$    & $s_2$           &  9:51  {~~~~~~~}                    & $s_3$           & 10:50   {~~~~~}                  &  9.83   {~~~~~~~~~~~~~~~~}                \\
  $c_7$    & $s_1$           & 10:12  {~~~~~~~}                    & $s_2$           & 10:46   {~~~~~}                  &  5.67   {~~~~~~~~~~~~~~~~}                \\ \toprule
  \end{tabular}}
\end{table}

Each vehicle's assignment plan is shown in~\Cref{fig:IllustrativeExample} as a sequence of arrows.~The rental and return events are represented by nodes that display pie charts visualizing the vehicle's remaining energy; the waiting order is displayed as a dashed arrow~(with a plugin symbol, if it is recharging) when it is available for rental; and the rental order as a full arrow with~a symbol bearing the customer's identification when it is fulfilling a customer's driving demand.

\begin{figure}[H]
  \centering
  \resizebox{\textwidth}{!}{%
    \input{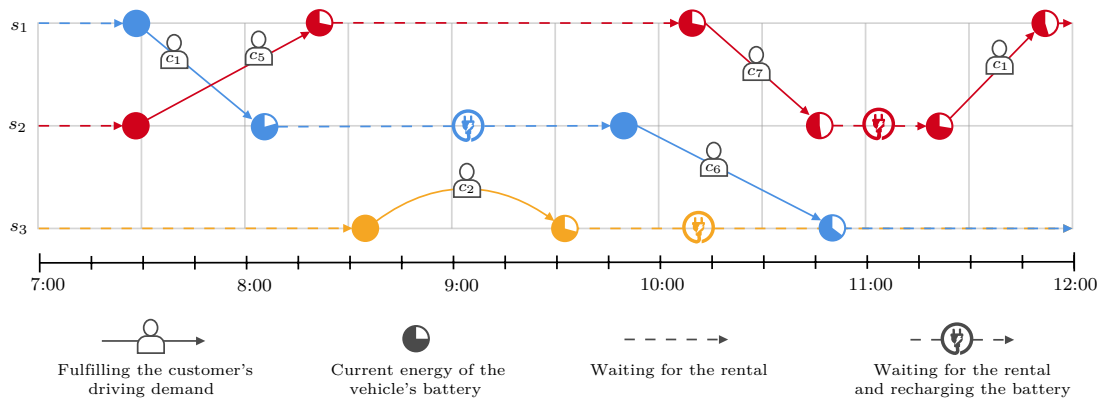}
  }
  \vspace*{-2mm}
  \caption{Illustration of assignment plans (visualized as a blue, red, and yellow sequence of arrows) for the rental requests in~\Cref{tab:reservation}.}
  \label{fig:IllustrativeExample}
\end{figure}

The assignment plans illustrated in~\Cref{fig:IllustrativeExample} show that customers~\(c_3\) and~\(c_4\) are turned away~(meaning that none of their demands have been fulfilled), whereas~\(c_1\), \(c_2\), \(c_5\), \(c_6\), and~\(c_7\) are served.~It is worth noting that~\(c_1\) has both of their demands fulfilled by different vehicles.~Despite being used for illustration only, those assignment plans represent an optimal solution.~Additionally, by rearranging the parking spaces where the vehicles are parked, or by choosing whether to charge them or not, several optimal solutions can be obtained.

\begin{remark}
  \normalfont When a driving demand is selected to be fulfilled by a vehicle~\(v\) from station~\(\outgoing{s}\) to~\(\incoming{s}\), starting at time~\(t_{i}\) and ending at time~\(t_{j}\), it means that the vehicle~\(v\) will be no longer parked in a parking space of the station~\(\outgoing{s}\) at time~\(t_{i}\) and that it will be parked in a parking space of the station~\(\incoming{s}\) at time~\(t_{j}\), in such a way that the fulfilling period means a left-closed and right-open interval.
\end{remark}

In addition, to specify the initial distribution of vehicles, we denote by~\mbox{\(\matheuler{V}^{\prime}_{s} \subseteq \matheuler{V}_{s}\)} the set of vehicles originally placed in the parking spaces equipped with charging facilities.~If the required energy~\(\varepsilon\) of the demand is not necessary, we omit it from the notation and denote the demand as a quadruple~\mbox{\((\outgoing{s}, t_{i}, \incoming{s}, t_{j} ) \in \matheuler{D}\)}.

\subsection{Our contributions and outline}

In this paper, we expand on the findings of~\cite{Silva/2023} with respect to the complexity, formulations, and exact solution of the~\shortname.~More specifically, we show theoretical results on the inapproximability, a search space analysis, a new modeling approach for the~\shortname, which is a two-index energy flow formulation —~unlike the usual three-index vehicle flow formulations commonly utilized in the literature~\citep{Kek/2009, Jorge/2015, Gambella/2018}.~The main contributions of this paper can be summarized as follows.

\begin{enumerate}[label=({\roman*})]
  \item A prove that, for any~\mbox{\(\epsilon > 0\)}, it is impossible to approximate~\shortname in polynomial time within a factor of~\(n^{1-\epsilon}\)\!, where~\(n\) is the number of customers, unless~\mbox{P = NP}.
  \item A novel mathematical formulation for the~\shortname, where the energy of each vehicle is used as continuous flow instead of binary flow in space-time networks.
  \item A relax-and-fix heuristic to produce high-quality solutions.
  \item An exact algorithm, which employs two schemes of variable-fixing and a warm-start solution to efficiently solve instances with several vehicles and parking spaces. 
  \item Optimal solutions are obtained for all instances of the previous benchmark.
  \item A new benchmark is proposed to raise the difficulty bar for future methodologies.
\end{enumerate}

The remainder of this paper is organized as follows.~In~\Cref{sec:approximability}, we present theoretical results on the inapproximability of the~\shortname and briefly address how it does not have a monotone structure that preserves the feasibility of a bottom-up (or top-down) search.~In~\Cref{sec:formulation3}, we describe our proposed mixed-integer linear programming formulation based on energy flows.~In~\Cref{sec:approaches}, we present a heuristic and an exact approach derived from our formulation.~In~\Cref{sec:experiments}, we present and discuss the computational results.~Finally, in~\Cref{sec:conclusions}, we give concluding remarks and an outlook on future research.

\section{Problem's complexity}
\label{sec:approximability}

In this section, we study the computational complexity of the~\shortname in a particular case where recharging vehicles is not necessary.~Even with this simplification, we can already obtain some unfavorable complexity results.

\subsection{Inapproximability results}

In this section, we present theoretical findings on the inapproximability of the~\shortname, including a proof that it is not in~APX, or simply put, it does not have a polynomial-time algorithm with constant approximation factor, unless~\mbox{P = NP}.~We go even further by showing it is not approximable by a factor of~\(n^{1-\epsilon}\), for any~\(\epsilon > 0\), unless~\mbox{P = NP}.

If a problem is strongly NP-hard, then it does not have a pseudo-polynomial time algorithm and it also does not have a fully-polynomial time approximation scheme~(FPTAS), unless~\mbox{P $=$ NP}~\citep{Vazirani/2001}.~\cite{Silva/2023} shown a polynomial reduction from the one-dimensional bin packing problem~(BPP), a well-known strongly NP-hard problem, to the~\shortname.~As a consequence, we have the following theorem.

\begin{corollary}[\cite{Silva/2023}, Corollary 4]
  \label{corollary:NP-hard}
  The~{\normalfont \shortname}\! is~\mbox{{\normalfont NP}-hard} in the strong sense even if every customer has only one demand and when there is only one station.
\end{corollary}

\cite{Bohmova/2016} have shown that the~TFC problem, a special case of the~\shortname without batteries, is~APX-hard using an~\mbox{L-reduction} from the~\mbox{\textsc{Max}-3SAT(3)}.

\begin{corollary}[\cite{Bohmova/2016}, Theorem 2]
  The {\normalfont\shortname}\! is {\normalfont APX}\!-hard even when there are only two stations, every customer has two demands, the rental time is the same for every demand, and there is only one vehicle without battery.
\end{corollary}

On the approximability of the~\shortname, an interesting question is whether there is a polynomial-time algorithm that guarantees a constant approximation factor.~We answer this question by showing that the~\shortname is not in~APX, unless~\mbox{P = NP}.~To this end, we propose a reduction from the maximum independent set problem, for which it is known that no polynomial-time algorithm can approximate it to within~\(n^{1-\epsilon}\) in general graphs with~\(n\) vertices, for any~\(\epsilon > 0\), unless~\mbox{P = NP}~\citep{Hastad/1999}.

{\noindent{\textbf{Maximum Independent Set Problem~(MISP).}}\it~Given a graph~\(\matheuler{G}=(\matheuler{U}, \matheuler{E})\), find an independent set in~\(\matheuler{G}\) of maximum cardinality, that is, a largest subset of vertices \(\matheuler{U}^{*} \subseteq \matheuler{U}\), such that there is no edge in~\(\matheuler{E}\) connecting two vertices in~\(\matheuler{U}^{*}\).}

\begin{theorem}
  \label{theorem:notAPX}
  There is a polynomial-time reduction from~{\normalfont MISP} to~{\normalfont\shortname}.
\end{theorem}

First we describe a construction and some properties used to prove~\Cref{theorem:notAPX}.

\noindent{\it\textBF{Instance Construction.}}~Let~\(\matheuler{I}_{\textsf{MISP}}\) denote an arbitrary instance of the~MISP given by a set of vertices \mbox{\(\matheuler{U} = \{u_1, \dots, u_n\}\)} and a set of edges~\(\matheuler{E} = \{e_1, \dots, e_m\}\).~The trick is to construct an instance of the~\shortname, denoted by~\(\matheuler{I}_{\textsf{\shortname}}\), from~\(\matheuler{I}_{\textsf{MISP}}\) in such a way that there is a one-to-one correspondence between the customers in the constructed instance and the vertices in the instance of the~MISP.~More precisely, each vertex~\(u \in \matheuler{U}\) is associated with a customer, denoted by~\(c_u\).~There is only one station and a single vehicle with enough charge to fulfill all the driving demands in the system.

In the following, we describe how the demands of the customers are organized into gadgets.~For every edge~\(e \in \matheuler{E}\), there is a gadget~\(\textsf{G}_e\) with more than one driving demand starting and ending in it, but configured so that all demands depart the station at once to ensure that the rental periods overlap~—~meaning that at most one demand placed in~\(\textsf{G}_e\) can be fulfilled.~The gadgets are placed sequentially.~See~\Cref{fig:MISPtoEVSP} for the exact configuration.

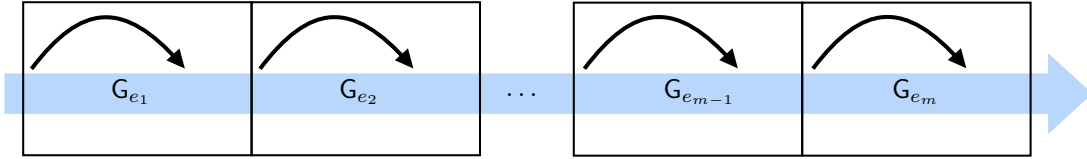
\begin{figure}[H]
  \centering
  \resizebox{\textwidth}{!}{%
    \tikzset{every picture/.style={line width=0.75pt}}

\begin{tikzpicture}[x=0.75pt, y=0.75pt, yscale=-1, xscale=0.96]

    \draw  [draw opacity=0][fill={rgb, 255:red, 185; green, 215; blue, 252 }  ,fill opacity=1 ] (27.53,73.89) -- (560.51,73.89) -- (560.51,63.89) -- (583.19,83.89) -- (560.51,103.89) -- (560.51,93.89) -- (27.53,93.89) -- cycle ;
    \draw [line width=1.5]    (41.33,71.67) .. controls (66.68,39.33) and (88.23,36.94) .. (117.41,68.97) ;
    \draw [shift={(119.67,71.5)}, rotate = 228.81] [fill={rgb, 255:red, 0; green, 0; blue, 0 }  ][line width=0.08]  [draw opacity=0] (9.29,-4.46) -- (0,0) -- (9.29,4.46) -- cycle    ;

    \draw [line width=1.5]    (158.03,71.62) .. controls (183.38,39.28) and (204.92,36.89) .. (234.1,68.92) ;
    \draw [shift={(236.36,71.45)}, rotate = 228.81] [fill={rgb, 255:red, 0; green, 0; blue, 0 }  ][line width=0.08]  [draw opacity=0] (9.29,-4.46) -- (0,0) -- (9.29,4.46) -- cycle    ;

    \draw [line width=1.5]    (323.73,71.57) .. controls (349.08,39.23) and (370.63,36.84) .. (399.81,68.87) ;
    \draw [shift={(402.07,71.4)}, rotate = 228.81] [fill={rgb, 255:red, 0; green, 0; blue, 0 }  ][line width=0.08]  [draw opacity=0] (9.29,-4.46) -- (0,0) -- (9.29,4.46) -- cycle    ;

    \draw [line width=1.5]    (440.43,71.52) .. controls (465.78,39.18) and (487.32,36.79) .. (516.5,68.82) ;
    \draw [shift={(518.76,71.35)}, rotate = 228.81] [fill={rgb, 255:red, 0; green, 0; blue, 0 }  ][line width=0.08]  [draw opacity=0] (9.29,-4.46) -- (0,0) -- (9.29,4.46) -- cycle    ;

    \draw (281.93,82.07) node [anchor=north west][inner sep=0.75pt]   [align=left] {$\displaystyle \dotsc $};
    \draw (479.31,75.07) node [anchor=north west][inner sep=0.75pt]   [align=left] {\(\textsf{G}_{e_m}\)};
    \draw (362.61,75.07) node [anchor=north west][inner sep=0.75pt]   [align=left] {\(\textsf{G}_{e_{m-1}}\)};
    \draw (196.91,75.07) node [anchor=north west][inner sep=0.75pt]   [align=left] {\(\textsf{G}_{e_2}\)};
    \draw (80.21,75.07) node [anchor=north west][inner sep=0.75pt]   [align=left] {\(\textsf{G}_{e_1}\)};

    \draw   (37.08,38.73) -- (153.73,38.73) -- (153.73,114.33) -- (37.08,114.33) -- cycle ;
    \draw   (153.73,38.73) -- (270.38,38.73) -- (270.38,114.33) -- (153.73,114.33) -- cycle ;
    \draw   (318.28,38.93) -- (434.93,38.93) -- (434.93,114.53) -- (318.28,114.53) -- cycle ;
    \draw   (434.93,38.93) -- (551.58,38.93) -- (551.58,114.53) -- (434.93,114.53) -- cycle ;

\end{tikzpicture}
  }
  \caption[Placement of the gadgets for the given instance of the~MISP]{Placement of the gadgets for the given instance of the~MISP.~For simplicity, there is only one bolded arrow in each gadget and it represents all its driving demands.}
  \label{fig:MISPtoEVSP}
\end{figure}

Let~\(\matheuler{N}_{u} = \{u\lowprime \in \matheuler{U} : (u, u\lowprime) \in \matheuler{E}\}\) be the neighborhood of a vertex~\(u \in \matheuler{U}\).~For each edge~\(e = (u, u\lowprime) \in \matheuler{E}\), customers~\(c_u\) and~\(c_{u^\prime}\) have a demand placed in~\(\textsf{G}_e\) with the rental periods equal to~\(1/|\matheuler{N}_{u}|\) and~\(1/|\matheuler{N}_{u^\prime}|\), respectively.~Note that since any customer~\(c_u\) has exactly~\(|\matheuler{N}_{u}|\) demands, the total rental time of their demands is equal to~\(1\).

Let~\(\rho\) be the polynomial time transformation from a~MISP instance to a~\shortname instance described above, such that~\mbox{\(\matheuler{I}_{\textsf{\shortname}} = \rho(\matheuler{I}_{\textsf{MISP}})\)}.~The properties of that transformation are described in the following.

\begin{corollary}
  \label{corollary:same_value}
  Let~{\normalfont val}\((\matheuler{S})\) denote the value of a solution \(\matheuler{S}\).~Let~\(\matheuler{S}_{\normalfont \textsf{\shortname}}\)\! denote a solution for~\mbox{\(\matheuler{I}_{\normalfont \textsf{\shortname}} = \rho(\matheuler{I}_{\normalfont \textsf{MISP}})\)}.~If a solution for~\(\matheuler{I}_{\normalfont \textsf{MISP}}\), denoted by~\(\matheuler{S}_{\normalfont \textsf{MISP}}\), is constructed from~\(\matheuler{S}_{\normalfont \textsf{\shortname}}\), then~\mbox{{\normalfont val}\((\matheuler{S}_{\normalfont \textsf{\shortname}}) = \)~{\normalfont val}\((\matheuler{S}_{\normalfont \textsf{MISP}})\)}.
\end{corollary}

{\noindent {\it\textBF{Proof.}}~The proof follows directly from the construction of~\(\matheuler{I}_{\normalfont \textsf{\shortname}}\).~We can use the fact that~whenever a customer~\(c_u\) is served, then any customer~\(c_{u^\prime}\), for~\(u\lowprime \in \matheuler{N}_{u}\), cannot be served.~\hfill \(\square\)
}

The above lemma immediately gives us the following corollary.

\begin{corollary}
  \label{corollary:same_value}
  Let~{\normalfont val}\((\matheuler{I}_{\normalfont \textsf{\shortname}})\)\! and~{\normalfont val}\((\matheuler{I}_{\normalfont \textsf{MISP}})\)\! denote, respectively, the value for the instances~\(\matheuler{I}_{\normalfont \textsf{\shortname}}\)\! and~\(\matheuler{I}_{\normalfont \textsf{MISP}}\), where~\mbox{\(\matheuler{I}_{\normalfont \textsf{\shortname}} = \rho(\matheuler{I}_{\normalfont \textsf{MISP}})\)}.~Then~\mbox{{\normalfont val}\((\matheuler{I}_{\normalfont \textsf{\shortname}}) = \)~{\normalfont val}\((\matheuler{I}_{\normalfont \textsf{MISP}})\)}.
\end{corollary}

{\noindent {\it\textBF{Proof {\normalfont (}of~\Cref{theorem:notAPX}{\normalfont )}.}}~It is an immediate consequence of~\Cref{lemma:same_solution} and~\Cref{corollary:same_value}.~\hfill \(\square\)
}

From~\Cref{theorem:notAPX}, we immediately deduce that the~\shortname is not in~APX, unless~\mbox{P = NP}, and we derive the following result. 

\begin{corollary}
  For any fixed~\mbox{\(\epsilon > 0\)}, the~{\normalfont\shortname}\! cannot be approximated in polynomial time within a factor of~\(n^{1-\epsilon}\)\!, where~\(n\) denotes the number of customers, unless {\normalfont P} \(=\) {\normalfont NP}, even if there is only one station and one vehicle.
\end{corollary}

\subsection{Non-monotone structure of the search space}
\label{sec:search_spaces}

In the face of the unfavorable results on the complexity and inapproximability of the~\shortname, we can assume that these results provide a theoretical justification for the employment of heuristics, which are the most popular approaches to tackling problems like this one.~However, as we shall discuss next, developing effective heuristics for the~\shortname is not an easy task.

In many combinatorial optimization problems, such as the \emph{knapsack problem}, any feasible and consequently optimal solution can be found by starting with an empty solution and iteratively adding elements~(\eg, items), one by one, in a predetermined order, without making the current solution infeasible.~In this case, we say that the optimization problem has an~\emph{upward-monotone structure}.~Unfortunately, for the~\shortname there are instances in which this property does not hold, taking into account any ordering of the customers.

In a bottom-up search, whenever a new customer is added to a given feasible solution~(or, in a top-down search, after a served customer is removed), a significant number of assignment plans may need to be reconfigured to maintain feasibility.~Any modification to an assignment plan, including a repair step, may have a cascading impact that leaves many customers as partially served, making the current solution infeasible.~In the worst case, it will be necessary to create a new assignment plan for each vehicle.


\Cref{fig:TFC-example2} gives a small example showing that the~\shortname does not have a monotone structure that preserves the feasibility of a bottom-up (or top-down) search.~We first notice that all four depicted customers can be served simultaneously, assuming that there is initially one vehicle available at each station with enough charge to fulfill all the driving demands in the system.~However, there is no feasible solution where exactly three customers are served.

\begin{figure}[H]
  \centering
  \resizebox{\textwidth}{!}{%
    \input{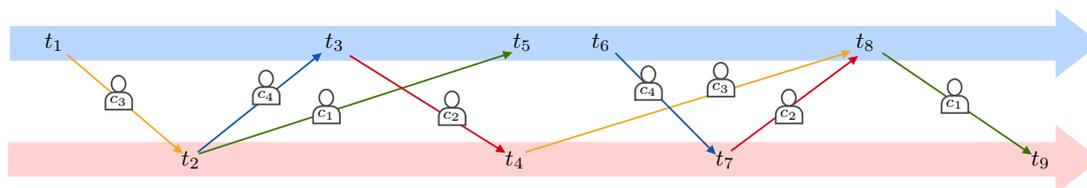}
  }
  \caption[An example where the upward-monotone structure does not hold]{An example where the upward-monotone structure does not hold.}
  \label{fig:TFC-example2}
\end{figure}

All theoretical results presented in this section have ignored the possibility of recharging the vehicle's battery.~This means that considering the energy consumption and recharging of vehicle batteries can further increase the complexity of developing efficient algorithms.

\section{Mathematical formulation}
\label{sec:formulation3}

In this novel formulation for the~\shortname, we abstract the explicit assignment of vehicles to demands, and their energy is modeled as flows in a space-time network, where the space dimension is represented by parking spaces.~When a demand is selected to be fulfilled, it is modeled as a flow of energy from the parking space~(origin node) at the pick-up station to the parking space~(destination node) at the drop-off station, with a reduction according to the energy consumed.~Further details are presented and discussed in the following paragraphs.

Let~\(\matheuler{P}_{s}\) be the set of parking spaces at station~\(s\), where~\mbox{\(|\matheuler{P}_{s}| = \textsf{C}_{s}\)}, for each~\mbox{\(s \in \matheuler{S}\)}.~Moreover, for each station~\mbox{\(s \in \matheuler{S}\)} and each parking space~\mbox{\(\scalebox{0.93}{$p$} \in \matheuler{P}_{s}\)}, let~\mbox{\(\matheuler{T}^{p}_{\vphantom{0}} = \{ t \in \matheuler{T} : \exists (s, t) \in \Pi \}\)} be the set of time instants for~\(\scalebox{0.93}{$p$}\), where~\(\Pi\) is the union of all sets~\mbox{\(\{ (\outgoing{s}, t_{i}), (\incoming{s}, t_{j}) \}\)}, for each~\mbox{\((\outgoing{s}, t_{i}, \incoming{s}, t_{j}) \in \matheuler{D}\)}, and let~\mbox{\(\matheuler{T}^{p}_{\,0} = \matheuler{T}^{p}_{\vphantom{0}} \cup \{t_{0}\}\)} be an extension of the set of time instants, where time~\(t_{0}\) indicates the initial configuration of the system.~We denote by~\mbox{\(\matheuler{P} = \bigcup_{s \in \matheuler{S}} \matheuler{P}_{s}\)} the set of all parking spaces.

We define a pair of decision variables~\(y^{p}_{t}\) and~\(\ell^{p}_{t}\), for all~\mbox{\(t \in \matheuler{T}^{p}_{\,0}\)}, where~\mbox{\(\scalebox{0.93}{$p$} \in \matheuler{P}_{s}\)}, for~\mbox{\(s \in \matheuler{S}\)}.~Each variable~\(y^{p}_{t}\) is binary and takes the value~\(1\) if and only if the parking space~\(\scalebox{0.93}{$p$}\) is occupied by a vehicle at time~\(t\); whereas each variable~\(\ell^{p}_{t}\) is continuous and stores the energy of the vehicle parked on~\(\scalebox{0.93}{$p$}\) at time~\(t\), such that the energy stored in~\(\ell^{p}_{t}\) is at most~\(\textsf{L}y^{p}_{t}\).

Whenever a demand is fulfilled, an origin parking space must transfer all stored energy in the form of a flow to a destination parking space, as explained earlier.~We define two sets of decision variables, for each~\mbox{\(d = (\outgoing{s}, t_{i}, \incoming{s}, t_{j}) \in \matheuler{D}\)}.~One of these sets is obtained~by defining a continuous variable~\(\outgoing{\ell}^{\:\!p}_{\hspace*{-.2mm}d}\), for each~\mbox{\(\scalebox{0.93}{$p$} \in \matheuler{P}_{\outgoing{s}}\)}; the other set is obtained by defining a continuous variable~\(\incoming{\ell}^{\:\!p}_{\hspace*{-.2mm}d}\), for each~\mbox{\(\scalebox{0.93}{$p$} \in \matheuler{P}_{\incoming{s}}\)}.~For example, a parking space~\(\scalebox{0.93}{$p$}\) flows energy at time~\(t\) to fulfill a demand~\mbox{\(d = (\outgoing{s}, t_{i}, \incoming{s}, t_{j}, \varepsilon)\)} only if~\mbox{\(\scalebox{0.93}{$p$} \in \matheuler{P}_{\outgoing{s}}\)}, \mbox{\(t_{i} = t\)} and~\mbox{\(\ell^{p}_{t-1} \geq \varepsilon\)}, leading to~\mbox{\(\outgoing{\ell}^{\:\!p}_{\hspace*{-.2mm}d} = \ell^{p}_{t-1}\)} and~\mbox{\(\ell^{p}_{t} = 0\)}.

We introduce the recharging processes as follows.~Let~\(\textsf{E}^{p}_{\,t}\) denote the amount of energy charged at parking space~\(\scalebox{0.93}{$p$}\) at time~\mbox{\(t \in \matheuler{T}^{p}_{\vphantom{0}}\)}, such that it is equal to~\mbox{\(\mu(t - t^{\prime})\)}, for~\mbox{\(t^{\prime} = \max \{ t^{\prime}\! \in \matheuler{T}^{p}_{\,0}\!: t^{\prime}\! < t\}\)}, if~\(p\) is equipped with a charging facility, and~\(0\) otherwise, for any parking space~\mbox{\(\scalebox{0.93}{$p$} \in \matheuler{P}\)}.

Figure~\ref{fig:energyModeling} illustrates the modeling of the energy flow for a parking space~\(\scalebox{0.93}{$p$}\) at time~\(t\), which is considered a node.~For such node, the incoming energy flow is either from a vehicle parked at the same parking space~\(p\) in time~\(t-1\) (represented by variable~\(\ell^{p}_{t-1}\)) or from a fulfilled demand~\(d\) with parking space~\(p\) as destination at time~\(t\) (represented as the sum of variables~\(\incoming{\ell}^{\:\!p}_{\hspace*{-.2mm}d}\)).~The incoming flow~\(\textsf{E}^{p}_{\,t}\) represents the battery recharge at parking space~\(p\) and time~\(t\).~The outgoing flows are analogous.

\begin{figure}[H]
   \centering
   \tikzset{every picture/.style={line width=0.75pt}}

\begin{tikzpicture}[x=0.75pt,y=0.75pt,yscale=-1,xscale=1]

   \draw  [color={rgb, 255:red, 39; green, 127; blue, 232 }  ,draw opacity=1 ] (240,152.96) .. controls (240,136.9) and (253.7,123.87) .. (270.59,123.87) .. controls (287.48,123.87) and (301.18,136.9) .. (301.18,152.96) .. controls (301.18,169.03) and (287.48,182.06) .. (270.59,182.06) .. controls (253.7,182.06) and (240,169.03) .. (240,152.96) -- cycle ;
   
   \draw [color={rgb, 255:red, 39; green, 127; blue, 232 }  ,draw opacity=1 ]   (213.18,94.93) -- (247.31,129.32) ;
   \draw [shift={(249.43,131.45)}, rotate = 225.21] [fill={rgb, 255:red, 39; green, 127; blue, 232 }  ,fill opacity=1 ][line width=0.08]  [draw opacity=0] (8.93,-4.29) -- (0,0) -- (8.93,4.29) -- cycle    ;
   \draw [color={rgb, 255:red, 39; green, 127; blue, 232 }  ,draw opacity=1 ]   (291.47,174.04) -- (325.89,208.87) ;
   \draw [shift={(328,211)}, rotate = 225.33] [fill={rgb, 255:red, 39; green, 127; blue, 232 }  ,fill opacity=1 ][line width=0.08]  [draw opacity=0] (8.93,-4.29) -- (0,0) -- (8.93,4.29) -- cycle    ;
   \draw [color={rgb, 255:red, 39; green, 127; blue, 232 }  ,draw opacity=1 ]   (213.18,211) -- (247.23,176.63) ;
   \draw [shift={(249.34,174.5)}, rotate = 134.73] [fill={rgb, 255:red, 39; green, 127; blue, 232 }  ,fill opacity=1 ][line width=0.08]  [draw opacity=0] (8.93,-4.29) -- (0,0) -- (8.93,4.29) -- cycle    ;
   \draw [color={rgb, 255:red, 39; green, 127; blue, 232 }  ,draw opacity=1 ]   (291.59,131.63) -- (325.53,97.23) ;
   \draw [shift={(327.64,95.1)}, rotate = 134.62] [fill={rgb, 255:red, 39; green, 127; blue, 232 }  ,fill opacity=1 ][line width=0.08]  [draw opacity=0] (8.93,-4.29) -- (0,0) -- (8.93,4.29) -- cycle    ;
   \draw [color={rgb, 255:red, 39; green, 127; blue, 232 }  ,draw opacity=1 ]   (270.39,109.89) -- (270.5,120.6) ;
   \draw [shift={(270.53,123.6)}, rotate = 269.41] [fill={rgb, 255:red, 39; green, 127; blue, 232 }  ,fill opacity=1 ][line width=0.08]  [draw opacity=0] (8.93,-4.29) -- (0,0) -- (8.93,4.29) -- cycle    ;
   \draw [color={rgb, 255:red, 39; green, 127; blue, 232 }  ,draw opacity=1 ]   (270.11,95.84) .. controls (282.88,105.06) and (254.82,99.85) .. (270.45,110.16) ;

   \draw (198.4,76.8) node [anchor=north west][inner sep=0.75pt]   [align=left] {$\ell^{p}_{t-1}$};
   \draw (263.4,78.2) node [anchor=north west][inner sep=0.75pt]   [align=left] {$\textsf{E}^{p}_{\,t}$};
   \draw (320.2,77) node [anchor=north west][inner sep=0.75pt]   [align=left] {\,$\ell^{p}_{t}$};
   \draw (192.8,212) node [anchor=north west][inner sep=0.75pt]   [align=left] {\,${\displaystyle \sum_{d}} \,\incoming{\ell}^{p}_{d}$};
   \draw (308.8,212) node [anchor=north west][inner sep=0.75pt]   [align=left] {\,${\displaystyle \sum_{d}} \,\outgoing{\ell}^{\:\!p}_{d}$};
   \draw (260.16,147.71) node [anchor=north west][inner sep=0.75pt]   [align=left] {\( \scalebox{0.93}{$p$}\hspace*{-.5mm}, t \)};

\end{tikzpicture}
   \vspace*{-2mm}
   \caption{An illustrative example of the incoming and outgoing energy flows.}
   \label{fig:energyModeling}
\end{figure}
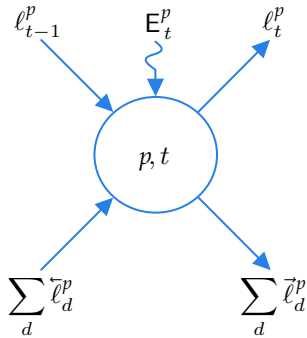

We define two additional sets of decision variables that indicate when energy flows over a demand in a parking space~\(\scalebox{0.93}{$p$}\) at time~\mbox{\(t \in \matheuler{T}^{p}\)}.~We obtain one of these sets by defining a binary variable~\(\outgoing{z}^{\:\!p}_{t}\) that takes value~\(1\) if there is a demand~\mbox{\(d = (\outgoing{s}, t_{i}, \incoming{s}, t_{j}) \in \matheuler{D}\)}, such that~\mbox{\(\scalebox{0.93}{$p$} \in \matheuler{P}_{\outgoing{s}}\)} and~\mbox{\(t_{i} = t\)}, and if and only if a demand~\(d\) is fulfilled by the vehicle parked on~\(\scalebox{0.93}{$p$}\).~Similarly, another set is obtained by defining a binary variable~\(\incoming{z}^{p}_{t}\) if there exists a demand~\mbox{\(d = (\outgoing{s}, t_{i}, \incoming{s}, t_{j}) \in \matheuler{D}\)}, such that~\mbox{\(\scalebox{0.93}{$p$} \in \matheuler{P}_{\incoming{s}}\)} and~\mbox{\(t_{j} = t\)}, which takes value~\(1\) if and only if any demand~\(d\) is fulfilled and the vehicle used for renting is parked at~\(\scalebox{0.93}{$p$}\).

We also define two other sets of decision variables indicating the movement of a vehicle out of and into a station when a demand~\mbox{\(d = (\outgoing{s}, t_{i}, \incoming{s}, t_{j})\)} is fulfilled, for each~\mbox{\(d \in \matheuler{D}\)}.~One is obtained by defining a binary variable~\(\outgoing{x}^{\:\!p}_{\hspace*{-.2mm}d}\), for each~\mbox{\(\scalebox{0.93}{$p$} \in \matheuler{P}_{\outgoing{s}}\)}, and the other by defining a binary variable~\(\incoming{x}^{p^{_\prime}}_{\hspace*{-.2mm}d}\)\!, for each~\mbox{\(\scalebox{0.93}{$p$}{\lowprime} \in \matheuler{P}_{\incoming{s}}\)}, that takes value~\(1\) if and only if~\(d\) is fulfilled by a vehicle in which it is picked-up at~\(\scalebox{0.93}{$p$}\) and dropped-off at~\(\scalebox{0.93}{$p$}{\lowprime}\), respectively.
~Finally, we define a binary variable~\(w_{c}\) that takes value~\(1\) if and only if customer~\mbox{\(c \in \matheuler{C}\)} is served.

The~\shortname is formulated as the following mixed-integer linear programming problem.

\noindent\RemoveSpaces{\makerefF{formulation:third}}
\begingroup
\allowdisplaybreaks
\small
\begin{alignat}{8}
\mbox{Maximize\,} & \sum\centerlap{c \in \matheuler{C}}\sum\rightlap{(%
                               \outgoing{s}\hspace*{-.3mm},%
                               t_{\hspace*{-.3mm}i}\hspace*{-.3mm},%
                               \incoming{s}\hspace*{-.3mm},%
                               t_{\hspace*{-.3mm}j}\!) \in \matheuler{D}_{c}\\} 
                \left( t_{j}-t_{i} \right) w_{c} & & & \makeref{EVSP3:ObjectiveFunction}\\[-7pt]
\mbox{Subject to}\! & \notag \\[-4pt]
            & \sum\rightlap{p \in \matheuler{P}_{\outgoing{s}}} \outgoing{x}^{\:\!p}_{\hspace*{-.2mm}d} = w_{c}
            & \forall c \in \matheuler{C},\,
              \forall d = (\outgoing{s}, t_{i}, \incoming{s}, t_{j}) \in \matheuler{D}_{c}
            &
            & \makeref{EVSP3:DemandConstraintOutgoing}\\
            & \sum\rightlap{p \in \matheuler{P}_{\incoming{s}}} \incoming{x}^{p}_{\hspace*{-.2mm}d} = w_{c}
            & \forall c \in \matheuler{C},\,
              \forall d = (\outgoing{s}, t_{i}, \incoming{s}, t_{j}) \in \matheuler{D}_{c}
            &
            & \makeref{EVSP3:DemandConstraintIncoming}\\
            & \sum\rightlap{d \in \matheuler{D}:
                s {\scriptscriptstyle =} \outgoing{s}\hspace*{-.2mm}, t {\scriptscriptstyle =} t_{i}
              }
              \outgoing{x}^{\:\!p}_{\hspace*{-.2mm}d} = \outgoing{z}^{\:\!p}_{t}
            & \forall s \in \matheuler{S},\,
              \forall \scalebox{0.93}{$p$} \in \matheuler{P}_{s},\,
              \forall t \in \matheuler{T}^{p}
            &
            & \makeref{EVSP3:ParkingSpotConstraintOutgoing}\\
            & \sum\rightlap{d \in \matheuler{D}:
                s {\scriptscriptstyle =} \incoming{s}\hspace*{-.2mm}, t {\scriptscriptstyle =} t_{j}
              }
              \incoming{x}^{p}_{\hspace*{-.2mm}d} = \incoming{z}^{p}_{t}
            & \forall s \in \matheuler{S},\,
              \forall \scalebox{0.93}{$p$} \in \matheuler{P}_{s},\,
              \forall t \in \matheuler{T}^{p}
            &
            & \makeref{EVSP3:ParkingSpotConstraintIncoming}\\            
            & ~\incoming{z}^{p}_{t} + y^{p}_{t^\prime} \leq 1
            & \forall \scalebox{0.93}{$p$} \in \matheuler{P},\,
              \forall t \in \matheuler{T}^{p}\!,\,
              t^{\prime}\! = \max \{t^{\prime}\! \in \matheuler{T}^{p}_{\,0}\!: t^{\prime}\! < t\}
            & 
            & \makeref{EVSP3:ParkingSpotConstraint}\\
            & ~y^{p}_{t} = y^{p}_{t^\prime} - \outgoing{z}^{\:\!p}_{t} + \incoming{z}^{p}_{t}
            & \forall \scalebox{0.93}{$p$} \in \matheuler{P},\,
              \forall t \in \matheuler{T}^{p}\!,\,
              t^{\prime}\! = \max \{t^{\prime}\! \in \matheuler{T}^{p}_{\,0}\!: t^{\prime}\! < t\}
            & 
            & \makeref{EVSP3:ParkingSpotConservation}\\
            & \sum\rightlap{p \in \matheuler{P}_{\outgoing{s}}} \outgoing{\ell}^{\,p}_{\hspace*{-.2mm}d} =
              \sum\rightlap{p \in \matheuler{P}_{\incoming{s}}} \incoming{\ell}^{\:\!p}_{\hspace*{-.2mm}d} +
                \varepsilon w_{c}
            & \forall d = (\outgoing{s}, t_{i}, \incoming{s}, t_{j}, \varepsilon) \in \matheuler{D}
            & 
            & \makeref{EVSP3:FlowConservationEquation}\\
            & ~\outgoing{\ell}^{\,p}_{\hspace*{-.2mm}d} \leq \textsf{L}\outgoing{x}^{\:\!p}_{\hspace*{-.2mm}d}
            & \forall d = (\outgoing{s}, t_{i}, \incoming{s}, t_{j}, \varepsilon) \in \matheuler{D},\,
              \forall \scalebox{0.93}{$p$} \in \matheuler{P}_{\outgoing{s}}
            & 
            & \makeref{EVSP3:ChargeLevelConstraints1}\\
            & ~\incoming{\ell}^{\:\!p}_{\hspace*{-.2mm}d} \leq (\textsf{L} - \varepsilon) \incoming{x}^{p}_{\hspace*{-.2mm}d}
            & \forall d = (\outgoing{s}, t_{i}, \incoming{s}, t_{j}, \varepsilon) \in \matheuler{D},\,
              \forall \scalebox{0.93}{$p$} \in \matheuler{P}_{\incoming{s}}
            & 
            & \makeref{EVSP3:ChargeLevelConstraints2}\\
            & \sum\rightlap{d \in \matheuler{D}:
                p \in \matheuler{P}_{\outgoing{s}}\!, t {\scriptscriptstyle =} t_{i}
              }
              \outgoing{\ell}^{\,p}_{\hspace*{-.2mm}d} + \ell^{p}_{t} \,\leq\, 
              \sum\rightlap{d \in \matheuler{D}:
                p \in \matheuler{P}_{\incoming{s}}\!, t {\scriptscriptstyle =} t_{j}
              }
              \incoming{\ell}^{\:\!p}_{\hspace*{-.2mm}d} + 
              \min (\ell^{p}_{t^{\prime}}\!+\!\textsf{E}^{p}_{\,t},\, \textsf{L} y^{p}_{t^{\prime}})
              \hspace*{-.77cm}
            & \forall \scalebox{0.93}{$p$} \in \matheuler{P},\,
              \forall t \in \matheuler{T}^{p}\!,\,
              t^{\prime}\! = \max \{t^{\prime}\! \in \matheuler{T}^{p}_{\,0}\!: t^{\prime}\! < t\}
            &
            & \makeref{EVSP3:RechargingConstraints}\\
            & ~y^{p}_{t_{0}} = 1,\, \ell^{p}_{t_{0}} = \textsf{L}^{v}
            & \scalebox{0.93}{$p$} = \psi(v),\,
              \forall v \in \matheuler{V}_{s},\,
              \forall s \in \matheuler{S}
            & 
            & \makeref{EVSP3:InitialDistribution} \\
            & ~y^{p}_{t_{0}} = 0,\, \ell^{p}_{t_{0}} = 0
            & \forall \scalebox{0.93}{$p$} \in \matheuler{P} 
              \!\setminus\! \{\psi(v) : \forall v \in \matheuler{V}_{s}\},\,
              \forall s \in \matheuler{S}
            & 
            & \makeref{EVSP3:InitialDistribution2}\\
            & ~y^{p}_{t} \in \{0, 1\},\, 
              \ell^{p}_{t} \in \mathbb{R}_{\geq 0}
            & \forall \scalebox{0.93}{$p$} \in \matheuler{P},\, 
              \forall t \in \matheuler{T}^{p}_{\,0}
            &
            & \makeref{EVSP3:Constraints1}\\
            & ~\outgoing{x}^{\:\!p}_{\hspace*{-.2mm}d} \in \{0, 1\},\, 
              \outgoing{\ell}^{\,p}_{\hspace*{-.2mm}d} \in \mathbb{R}_{\geq 0}
            & \forall d = (\outgoing{s}, t_{i}, \incoming{s}, t_{j}) \in \matheuler{D},\, 
              \forall \scalebox{0.93}{$p$} \in \matheuler{P}_{\outgoing{s}}
            &
            & \makeref{EVSP3:Constraints2}\\
            & ~\incoming{x}^{p}_{\hspace*{-.2mm}d} \in \{0, 1\},\, 
              \incoming{\ell}^{\:\!p}_{\hspace*{-.2mm}d} \in \mathbb{R}_{\geq 0}
            & \forall d = (\outgoing{s}, t_{i}, \incoming{s}, t_{j}) \in \matheuler{D},\, 
              \forall \scalebox{0.93}{$p$} \in \matheuler{P}_{\incoming{s}}
            &
            & \makeref{EVSP3:Constraints3}\\
            & ~\outgoing{z}^{\:\!p}_{t} \in \{0, 1\}
            & \forall \scalebox{0.93}{$p$}, t \in \matheuler{P} \!\times\! \matheuler{T}:
              \exists d = (\outgoing{s}, t_{i}, \incoming{s}, t_{j}) \in \matheuler{D},\, 
              \scalebox{0.93}{$p$} \in \matheuler{P}_{\outgoing{s}}\!, t = t_{i}
            &
            & \makeref{EVSP3:Constraints4}\\
            & ~\incoming{z}^{p}_{t} \in \{0, 1\}
            & \forall \scalebox{0.93}{$p$}, t \in \matheuler{P} \!\times\! \matheuler{T}:
              \exists d = (\outgoing{s}, t_{i}, \incoming{s}, t_{j}) \in \matheuler{D},\, 
              \scalebox{0.93}{$p$} \in \matheuler{P}_{\incoming{s}}\!, t = t_{j}
            &
            & \makeref{EVSP3:Constraints5}\\
            & ~w_{c} \in \{0, 1\}
            & \forall c \in \matheuler{C}
            &
            & \makeref{EVSP3:Constraints7}  
\end{alignat}
\endgroup

The objective function~\eqrefsmallsize{EVSP3:ObjectiveFunction} maximizes the sum of the rental time.~Constraint sets~\eqrefsmallsize{EVSP3:DemandConstraintOutgoing} and~\eqrefsmallsize{EVSP3:DemandConstraintIncoming} ensure that each fulfilled demand uses a pick-up and drop-off parking space, respectively, and that a customer is served if all of the customer's demands are fulfilled.~Similarly, constraint sets~\eqrefsmallsize{EVSP3:ParkingSpotConstraintOutgoing} and~\eqrefsmallsize{EVSP3:ParkingSpotConstraintIncoming}, respectively, impose that each parking space is free or occupied when the demand is fulfilled~(\ie, when there is an energy flow that fulfills a demand).~Constraint set~\eqrefsmallsize{EVSP3:ParkingSpotConstraint} imposes the criteria for a parking space to be occupied, and constraint set~\eqrefsmallsize{EVSP3:ParkingSpotConservation} ensures that the scheduling of parking spaces is feasible.~Note that to simplify the notation, we omit the existence conditions for the variables~\(\outgoing{z}^{\:\!p}_{t}\) and~\(\incoming{z}^{p}_{t}\) from the constraint sets~\eqrefsmallsize{EVSP3:ParkingSpotConstraintOutgoing} to~\eqrefsmallsize{EVSP3:ParkingSpotConservation}.

Furthermore, constraint sets~\eqrefsmallsize{EVSP3:FlowConservationEquation} to~\eqrefsmallsize{EVSP3:RechargingConstraints} represent the conservation of energy flows.~More specifically, set~\eqrefsmallsize{EVSP3:FlowConservationEquation} imposes that for each fulfilled demand, the outgoing energy flow from the pick-up parking space must be equal to the incoming energy flow at the drop-off parking space plus the energy consumed by the rental; whereas sets~\eqrefsmallsize{EVSP3:ChargeLevelConstraints1} and~\eqrefsmallsize{EVSP3:ChargeLevelConstraints2} limit the energy flow that fulfills a demand, and the set~\eqrefsmallsize{EVSP3:RechargingConstraints} limits the energy stored in each parking space for any given time instant to ensure the recharging process and prevent the battery capacity from being exceeded by charging.

Next, constraint sets~\eqrefsmallsize{EVSP3:InitialDistribution} to~\eqrefsmallsize{EVSP3:InitialDistribution2} specify an initial state of vehicles at each station, where~\(\psi\) is an injection function from~\(\matheuler{V}_{s}\) to~\(\matheuler{P}_{s}\), for each~\mbox{\(s \in \matheuler{S}\)}, that assigns each vehicle to a parking space, such that if a vehicle~\(v\) belongs to~\(\matheuler{V}^{\prime}_{s}\), then~\(v\) is associated with a parking space equipped with a charging facility~—~otherwise,~\(v\) is associated with a parking space that is not equipped with a charging facility.~Finally, expressions~\eqrefsmallsize{EVSP3:Constraints1} to~\eqrefsmallsize{EVSP3:Constraints7} are the domain constraints for the variables.

\begin{remark}
  \label{remark:evsp3-relaxed}
  \normalfont The sets of binary variables~\(w_{c}\), \(\outgoing{z}^{\:\!p}_{t}\) and~\(\incoming{z}^{p}_{t}\) can be relaxed to be continuous within the lower and upper binary bounds since these variables are guaranteed to take a binary value by constraint sets~\eqrefsmallsize{EVSP3:DemandConstraintOutgoing}~(and also~\eqrefsmallsize{EVSP3:DemandConstraintIncoming}), \eqrefsmallsize{EVSP3:ParkingSpotConstraintOutgoing} and~\eqrefsmallsize{EVSP3:ParkingSpotConstraintIncoming}, respectively.
\end{remark}

It is important to note that the set of binary variables~\(y^{p}_{t}\) can also be relaxed, and this can be proven by induction on the time instants, like the following.

\begin{proposition}
  \label{proposition:EVSP3_integer}
  The formulation~{\normalfont\ref{formulation:third}} has an integer-valued feasible solution even when the integrality constraint in~\eqrefsmallsize{EVSP3:Constraints1} is relaxed to be continuous.
\end{proposition}


{\noindent {\bf Proof.}~This proof is made by induction on~\(i \in \mathbb{Z}_{\geq 0}\), for~\(i \leq n\), here used to index the time instants in~\(\matheuler{T}^{p} = \{t_{0}, \dots, t_{n}\}\), for any~\(p \in \matheuler{P}\).~When~\(i = 0\) (\ie, for the initial time instant), the variable~\(y^{p}_{t_{0}}\) can only be~\(0\) or~\(1\), which follows directly from constraint sets~\eqrefsmallsize{EVSP3:InitialDistribution} and \eqrefsmallsize{EVSP3:InitialDistribution2}.~Let us assume, by induction, that the proposition holds for a given integer~\(k\), such that~\mbox{\(0 \leq k < n\)}, and consider the case for~\mbox{\(i = k+1\)}.~Then, taking into account the constraint set~\eqrefsmallsize{EVSP3:ParkingSpotConservation}, in which the equation can be rewritten as
\[y^{p}_{t_i} = y^{p}_{t_{i-1}} - \outgoing{z}^{\:\!p}_{t_i} + \incoming{z}^{p}_{t_i},\]
\noindent we will prove that the left-hand term can only be~\(0\) or~\(1\).

By the induction hypothesis,~we may assume that the variable~\(y^{p}_{t_{i-1}}\) takes a binary value; and the variables~\(\outgoing{z}^{\:\!p}_{t_i}\) and~\(\incoming{z}^{p}_{t_i}\) take only binary values, even relaxed to be continuous (see \Cref{remark:evsp3-relaxed}).~Therefore, the variable~\(y^{p}_{t_{i}}\) has an integer value.~Since~\(y^{p}_{t_{i}}\) is within the lower and upper binary bounds, it can only take a binary value.~This completes the proof.}~\hfill \(\square\)

\section{\ref*{formulation:third}-based solution approaches}
\label{sec:approaches}

In this section, we present two approaches for obtaining high-quality solutions and reducing the computational effort, both derived from the formulation~\mbox{\ref{formulation:third}}.~Our first approach is a relax-and-fix heuristic based on the idea of fixing the zero-valued variables in a relaxed solution to reduce the model size~—~a similar approach can be found in~\cite{Yee/1998}.~Our second approach is an exact methodology that employs a reduced-cost variable-fixing technique using the solution of the heuristic as a warm-start solution.

\subsection{Relax-and-fix heuristic}

We can observe in formulation~\mbox{\ref*{formulation:third}} that symmetries arise from the fact that the variables associated with the parking spaces, equipped or unequipped with a charging facility, located at the same station and time, can be permuted without changing the structure of the solution.~If we know ahead which variables are associated with the parking spaces with a low probability of appearing in an optimal solution, we can heuristically cut out part of the search space by fixing some of these variables to zero, consequently breaking part of the symmetry.~For this purpose, the knowledge regarding an optimal fractional solution from the linear programming relaxation can be exploited.

The concept behind this approach is simple.~We first relax the integrality constraints and solve the linear programming to obtain an optimal fractional solution.~From this solution, we fix to zero all the binary variables associated with the parking spaces whose values are zero —~they have a presumably low likelihood of appearing in a high-quality solution.~Then, by imposing the integrality constraints to obtain a feasible solution for the~\shortname, we optimally solve the resulting model.~Another implication of doing this is that the resulting model also has fewer constraints than the original one since those constraints consisting solely of the fixed variables are also inactive.~For a better understanding, the following pseudo-code summarizes this approach.

\begin{algorithm}[H]
  \caption{The linear relaxation-based variable-fixing (LRBVF\label{algo:EVSP3-H}) heuristic.}
  \begin{algorithmic}[1]
    \State Relax the integrality constraints of all binary variables in the~\ref*{formulation:third} model.
    \State Solve the relaxed model to obtain an optimal fractional solution.
    \State Fix to zero the variables~\(\outgoing{x}^{\:\!p}_{\hspace*{-.2mm}d}, \outgoing{z}^{\:\!p}_{t}, \incoming{x}^{p}_{\hspace*{-.2mm}d}\) and~\(\incoming{z}^{p}_{t}\) if their value in the solution is zero.
    \State Restore the integrality constraints for the variables~\(\outgoing{x}^{\:\!p}_{\hspace*{-.2mm}d}\) and \(\incoming{x}^{p}_{\hspace*{-.2mm}d}\).
    \State Solve the resulting model.
    \State Return the integer-valued solution achieved.
  \end{algorithmic}
\end{algorithm}


\begin{proposition}
  \label{proposition:LRBVF}
  The algorithm~{\normalfont\hyperref[algo:EVSP3-H]{LRBVF}} finds a feasible solution to any instance of the~{\normalfont\shortname}.
\end{proposition}

{\noindent {\bf Proof.}~Note first that, given~\Cref{remark:evsp3-relaxed} and~\Cref{proposition:EVSP3_integer}, only the integrality of the variables~\(\outgoing{x}^{\:\!p}_{\hspace*{-.2mm}d}\) and \(\incoming{x}^{p}_{\hspace*{-.2mm}d}\) is imposed (in step~\(4\)).~Since the algorithm only fixes variables related to fulfilling demands (in step~\(3\)), the vehicles are free at any time to remain parked in a parking space.~Thus, the~LRBVF algorithm ensures that a feasible solution is always found.}~\hfill \(\square\)


\subsection{Exact algorithm with reduced-cost variable-fixing}

An optimal solution can be achieved based on the observation that we can just extend the~\hyperref[algo:EVSP3-H]{LRBVF} heuristic by using its solution as a warm-starting point, which can be done by releasing the variables previously fixed and solving the updated model from the current solved state.~To introduce a better exact approach, we propose an algorithm that extends our heuristic by performing two additional variable-fixing schemes.

Our first scheme makes use of the reduced-cost variable-fixing technique, which can be attempted whenever a dual feasible solution and a lower bound corresponding to a feasible solution are available.~Consider the following \(0\)\hspace{.2mm}-\(1\) integer linear programming problem 
\leqnomode
\begin{equation}\tag{P}\label{eq:P}
  \max\bigl\{\boldsymbol{c}^{_\textsf{T}}\boldsymbol{x} \,:\, \boldsymbol{A}\boldsymbol{x} = \boldsymbol{b},~ \boldsymbol{x} \in \{0, 1\}^{n} \bigr\}{\normalfont ,}
\end{equation}
\noindent where~\(\boldsymbol{A} \in \mathbb{R}^{m \Times n}\), \(\boldsymbol{b} \in \mathbb{R}^{m}\), and~\(\boldsymbol{c} \in \mathbb{R}^{n}\), for some \(m \geq n \geq 1\).
~The dual problem associated with the linear relaxation of~P can be presented in standard form as
\begin{equation}\tag{D}\label{eq:D}
  \min\bigl\{\boldsymbol{y}^{_\textsf{T}}\boldsymbol{b} + \boldsymbol{u}^{_\textsf{T}}\boldsymbol{1} :\, \boldsymbol{y}^{_\textsf{T}}\!\boldsymbol{A} + \boldsymbol{u} - \boldsymbol{v} = \boldsymbol{c},~ \boldsymbol{u}, \boldsymbol{v} \geq \boldsymbol{0}\bigr\}{\normalfont ,}
\end{equation}
\reqnomode
\noindent where~\(\boldsymbol{y} \in \mathbb{R}^{m}\) is the vector of dual variables,~\(\boldsymbol{u},\boldsymbol{v} \in \mathbb{R}^{n}\) are the vectors of dual slack variables, and~\(\boldsymbol{0}\) and~\(\boldsymbol{1}\) are~\mbox{\(n\)-vectors} of zeros and ones, respectively.~Concerning a feasible dual solution~\((\bar{\boldsymbol{y}}, \bar{\boldsymbol{u}}, \bar{\boldsymbol{v}})\) of~D, by complementary slackness, the primal value~\(\boldsymbol{c}^{_\textsf{T}}\boldsymbol{x}\) can be written as
\begin{align}
  \boldsymbol{c}^{_\textsf{T}}\boldsymbol{x} 
  & = (\bar{\boldsymbol{y}}^{_\textsf{T}}\!\boldsymbol{A} + \bar{\boldsymbol{u}} - \bar{\boldsymbol{v}})^{_\textsf{T}}\boldsymbol{x}&\nonumber\\                                                                                   
  & = \bar{\boldsymbol{y}}^{_\textsf{T}}\boldsymbol{b} + \bar{\boldsymbol{u}}^{_\textsf{T}}\boldsymbol{x} - \bar{\boldsymbol{v}}^{_\textsf{T}}\boldsymbol{x}&\nonumber\\                                                                                                                              
  & = \scalebox{1.2}{$\bar{z}$} + \bar{\boldsymbol{u}}^{_\textsf{T}}(\boldsymbol{x}-\boldsymbol{1}) - \bar{\boldsymbol{v}}^{_\textsf{T}}\boldsymbol{x},&\nonumber
\end{align}
\noindent where~\(\scalebox{1.2}{$\bar{z}$}\) is the dual value from a feasible dual solution.~Let~\(\scalebox{1.2}{$\ubar{z}$}\) be a lower bound on the optimal value of~P.~Note that if~\(u_{i} > \scalebox{1.2}{$\bar{z}$}-\scalebox{1.2}{$\ubar{z}$}\), then~\(\boldsymbol{c}^{_\textsf{T}}\bar{\boldsymbol{x}} < \scalebox{1.2}{$\ubar{z}$}\) for any feasible solution~\(\bar{\boldsymbol{x}}\) with~\(\bar{x}_{i} = 0\), and if~\(v_{i} > \scalebox{1.2}{$\bar{z}$}-\scalebox{1.2}{$\ubar{z}$}\), then~\(\boldsymbol{c}^{_\textsf{T}}\bar{\boldsymbol{x}} < \scalebox{1.2}{$\ubar{z}$}\) for any feasible solution~\(\bar{\boldsymbol{x}}\) with~\(\bar{x}_{i} = 1\).~Hence, adapted for our purposes, we use the following theorem.

\begin{theorem}[\cite{Mitchell/1997}, Theorem 1] Let~\((\bar{\boldsymbol{y}}, \bar{\boldsymbol{u}}, \bar{\boldsymbol{v}})\) be a feasible solution of~D with value~\(\scalebox{1.1}{$\bar{z}$}\), and~\(\scalebox{1.1}{$\ubar{z}$}\) be a known lower bound on the optimal value of~{\normalfont P}.~Thus, the following two properties hold.
  \begin{enumerate}[(a)]
    \item If \(\bar{u}_{i} > \scalebox{1.2}{$\bar{z}$} - \scalebox{1.2}{$\ubar{z}$}\), then~\(x_{i}\) is equal to one in any optimal integer solution.
    \item If \(\bar{v}_{i} > \scalebox{1.2}{$\bar{z}$} - \scalebox{1.2}{$\ubar{z}$}\), then~\(x_{i}\) is equal to zero in any optimal integer solution.
  \end{enumerate}
\end{theorem}

Bearing in mind that the reduced cost~\(r_{i}\) of a variable~\(x_{i}\) is given by~\(r_{i} = c_{i} - \bar{\boldsymbol{y}}^{_\textsf{T}}\boldsymbol{A}_{^{\scalebox{1.2}{$.$}}i} = \bar{u}_{i}-\bar{v}_{i}\), where~\(\boldsymbol{A}_{^{\scalebox{1.2}{$.$}}i}\) denotes the~\(i\)-th column of~\(\boldsymbol{A}\), and that at most one of the dual slack variables~\(\bar{u}_i\) and~\(\bar{v}_i\) can be non-zero, we have the relations: when \(x_{i} = 1\), \(r_{i}\) is equal to~\(\bar{u}_i\), and when~\(x_{i} = 0\), \(r_{i}\) is equal to~\(-\bar{v}_i\).~As a result, the variable~\(x_{i}\) can be fixed at~\(1\) when~\(r_{i} > \scalebox{1.2}{$\bar{z}$} - \scalebox{1.2}{$\ubar{z}$}\) and at~\(0\) when~\(-r_{i} > \scalebox{1.2}{$\bar{z}$} - \scalebox{1.2}{$\ubar{z}$}\) without loss of optimality.~In our approach, we are only interested in fixing the variables~\(w_{c}\), which is useful to carry out the second variable-fixing scheme.

Our second scheme consists in fixing to zero variables related to the parking spaces.~Let~\mbox{\(\matheuler{C}^{\prime} \subseteq \matheuler{C}\)} denote the set of customers whose variables~\(w_{c}\), for~\(c \in \matheuler{C}^{\prime}\), have not been previously fixed to zero.~Additionally, for each~\(s \in \matheuler{S}\), let~\(\matheuler{P}^{\prime}_{s} = \{\scalebox{0.93}{$p$} \in \matheuler{P}_{s} : \scalebox{0.93}{$p$} \neq \psi(v), \forall v \in \matheuler{V}_{s}\}\) denote the set of parking spaces~(also called~\emph{absent parking spaces}) that are empty at the beginning.~By using the set~\(\matheuler{C}^{\prime}\), iteratively by the time instant, we can know the maximum amount of vehicles that can enter the station~\(s \in \matheuler{S}\) or, more specifically, the maximum amount of parking spaces that can be taken into account in any feasible solution, and then these parking spaces are taken off from~\(\matheuler{P}^{\prime}_{s}\).~Thus, the variables~\(y^{p}_{t}\), \(\outgoing{z}^{\:\!p}_{t}\), and~\(\incoming{z}^{p}_{t}\) associated with the remaining parking spaces in~\(\matheuler{P}^{\prime}_{s}\) can be fixed to zero.

\Cref{algo:subroutine} outlines the fixing of such variables, which can exclude some optimal solutions but are guaranteed not to eliminate all optimal solutions.~In this algorithm, we use the threshold~\(k\) as an upper bound on the total amount of parking spaces without charging facilities to be removed from the current set of remaining parking spaces~(\ie, the subset~\(\matheuler{U}\) of~\(\matheuler{P}^{\prime}_{s}\)).~When the initial value of~\(k\)~(step~\(4\)) is less than or equal to~\(0\), this means that all of the parking spaces without charging facilities will have their variables fixed at zero at any time instant.~Note that in the innermost loop (steps~\(5\) to~\(15\)), we look for the maximum amount of vehicles that can enter the station~\(s\) at time~\(t\) (step~\(6\)), to help determine the number of parking spaces to be removed from~\(\matheuler{E}\) and~\(\matheuler{U}\) (steps~\(9\) and~\(12\), respectively).~If there exist parking space~\(\scalebox{0.93}{$p$}\) in~\(\matheuler{E}\cup\matheuler{U}\), the value of~\(y^{p}_{t}\), \(\outgoing{z}^{\:\!p}_{t}\), and~\(\incoming{z}^{p}_{t}\) are then set to zero~(step~\(15\)).

\begin{algorithm}[H]
  \caption{The subroutine for fixing variables associated with the parking spaces.}\label{algo:subroutine}
  \begin{algorithmic}[1]
    \ForEach{station \(s\) in \(\matheuler{S}\)}
      \State Let \(\matheuler{D}^{\prime}_{s}\) denote the set of all~\((\outgoing{s}, t_{i}, \incoming{s}, t_{j}) \in \matheuler{D}_{c}\), such that~\(\outgoing{s} \neq \incoming{s}\) and~\(\incoming{s} = s\), for each~\(c \in \matheuler{C}^{\prime}\)\!.
      \State Let \(\matheuler{E} = \{\scalebox{0.93}{$p$} \in \matheuler{P}^{\prime}_{s} : \scalebox{0.93}{$p$} \text{ is equipped with a charging facility}\}\) and~\(\matheuler{U} = \matheuler{P}^{\prime}_{s}\!\setminus\!\matheuler{E}\) denote subsets of~\(\matheuler{P}^{\prime}_{s}\).
      \State Set the initial value of the threshold \(k\) equal to \(\min (|\matheuler{D}^{\prime}_{s}|-|\matheuler{E}|, |\matheuler{U}|)\).
      \ForEach{time \(t\) in \(\matheuler{T}\) (given in ascending order)}
      \State Let \(\textsf{M} = |\{(\outgoing{s}, t_{i}, \incoming{s}, t_{j}) \in \matheuler{D}^{\prime}_{s}\! : t_{j} = t\}|\) be the maximum number of vehicles at time~\(t\).
      \If{\(\textsf{M}\) is greater than zero}
        \If{\(\min (|\matheuler{E}|, \textsf{M})\) is greater than zero}
          \State Remove \(\min (|\matheuler{E}|, \textsf{M})\) parking spaces from \(\matheuler{E}\).
        \EndIf
        \If{\(k\) is greater than zero}
          \State Set \(n\) equal to \(\min (k, \textsf{M})\).
          \State Remove \(n\) parking spaces from~\(\matheuler{U}\).
          \State Decrease the threshold~\(k\) by~\(n\), \ie, \(k \gets k - n\).
        \EndIf
      \EndIf
      \If{\(\matheuler{E}\cup\matheuler{U}\) is not empty}
        \State Fix to zero the value of~\(y^{p}_{t}\), \(\outgoing{z}^{\:\!p}_{t}\) and~\(\incoming{z}^{p}_{t}\) for all remaining parking space~\(\scalebox{0.93}{$p$} \in \matheuler{E}\cup\matheuler{U}\).
      \EndIf
      \EndFor
    \EndFor
  \end{algorithmic}
\end{algorithm}

It should be noted that the maximum number of absent parking spaces from~\(\matheuler{P}^{\prime}_{s}\), for each station~\(s \in \matheuler{S}\), used in any optimal solution, is given by~\(\min(|\matheuler{P}^{\prime}_{s}|, |\matheuler{D}^{\prime}_{s}|)\).~To best explain~\Cref*{algo:subroutine}, keep in mind before the loop in step~\(5\) starts, \mbox{\(\min(|\matheuler{P}^{\prime}_{s}|, |\matheuler{D}^{\prime}_{s}|) = 2m + r\)}, where~\mbox{\(m = \min(k, |\matheuler{E}|)\)} is the number of times where parking spaces with and without charging facilities are taken into account simultaneously in the innermost loop of the algorithm.~More specifically, one parking space from~\(\matheuler{E}\) and~\(\matheuler{U}\) is removed for the first~\(m\) possible vehicle arrivals, and one parking space from~\(\matheuler{E}\) or~\(\matheuler{U}\) is removed for the last~\(r\) possible vehicle arrivals.

\begin{proposition}
  \label{proposition:RCBVF}
  Let~\mbox{\(\matheuler{C}^{\prime} \subseteq \matheuler{C}\)} be an arbitrary subset of customers.~If there is at least one optimal solution such that only the customers in~\(\matheuler{C}^{\prime}\) are served, then~{\normalfont\Cref{algo:subroutine}} does not eliminate all optimal solutions.
\end{proposition}

{\noindent {\bf Proof.}~It should be noted that the algorithm gives priority to using absent parking spaces with charging facilities.~Hence, we consider from now on that the optimal solutions make the most of the parking spaces with charging facilities.~Now, note that for a given iteration~\(s\) (step~\(1\) of~\Cref*{algo:subroutine}), considering each potential arrival of a vehicle from a station other than~\(s\), there are three cases:
\begin{enumerate}[label=({\roman*})]
  \item If the initial value of~\(k\) is less than or equal to~\(0\), then only parking spaces in~\(\matheuler{P}^{\prime}_{s}\) with charging facilities will be considered, and as long as~\(\matheuler{E}\) is not empty, one of these parking spaces is always chosen to be taken into account in a solution.
  \item If there are only parking spaces in~\(\matheuler{P}^{\prime}_{s}\) without charging facilities, then the initial value of the threshold~\(k\) is equal to \(\min (|\matheuler{D}^{\prime}_{s}|, |\matheuler{U}|)\), and as long as~\(\matheuler{U}\) is not empty, one of the parking spaces is always chosen to be taken into account in a solution.
  \item Otherwise, while~\(k\) and \(|\matheuler{E}|\) are positive values, one parking space with and another without a charging facility are concurrently chosen to be taken into account in a solution.~After that, is chosen to be considered at most one of the parking spaces from~\(\matheuler{U}\) (if~\(k\) is greater than~\(0\)) or~\(\matheuler{E}\) (if it is not empty).
\end{enumerate}

Without loss of generality, we assume that cases~(i) and~(ii) do not occur, and we will show that case~(iii) does not eliminate all optimal solutions.

Let~\(\matheuler{E}^{\scalebox{.63}{$($}\hspace{-.2mm}t\hspace{-.18mm}\scalebox{.63}{$)$}}\) and~\(\matheuler{U}^{\scalebox{.63}{$($}\hspace{-.2mm}t\hspace{-.18mm}\scalebox{.63}{$)$}}\) denote the sets of all the parking spaces that have been taken away from~\(\matheuler{E}\) and~\(\matheuler{U}\) in steps~\(9\) and~\(12\), respectively, up until the time~\(t\) along the innermost loop.~For each time~\(t\), by the pigeonhole principle, the larger value for the number of absent parking spaces with and without charging facilities that an optimal solution uses results in at most~\(|\matheuler{E}^{\scalebox{.63}{$($}\hspace{-.2mm}t\hspace{-.18mm}\scalebox{.63}{$)$}}\hspace{-.2mm}|\) and~\(|\matheuler{U}^{\scalebox{.63}{$($}\hspace{-.2mm}t\hspace{-.18mm}\scalebox{.63}{$)$}}\hspace{-.2mm}|\), respectively.~As a result, we can assure that, given a subset~\(\matheuler{C}^{\prime} \subseteq \matheuler{C}\) as stated,~\Cref*{algo:subroutine} will lead to at least an optimal solution.}~\hfill \(\square\)

Now that the variable-fixing schemes have been introduced, we present our exact approach, which is described in the following pseudo-code.

\begin{algorithm}[H]
  \caption{The reduced-cost-based variable-fixing (RCBVF\label{algo:EVSP3-E}) algorithm.}
  \algcomment{\emph{Note.} The vectors \(\bar{\boldsymbol{w}}\), \(\bar{\boldsymbol{x}}\), \(\bar{\boldsymbol{y}}\), \(\bar{\boldsymbol{z}}\), and~\(\bar{\boldsymbol{\ell}}\) denote the values of~\((w_{c})\), \((\outgoing{x}^{\:\!p}_{\hspace*{-.2mm}d}, \incoming{x}^{p}_{\hspace*{-.2mm}d})\), \((y^{p}_{t})\), \((\outgoing{z}^{\:\!p}_{t}, \incoming{z}^{p}_{t})\), and~\((\ell^{p}_{t}, \outgoing{\ell}^{\,p}_{\hspace*{-.2mm}d}, \incoming{\ell}^{\:\!p}_{\hspace*{-.2mm}d})\), respectively.}
  \begin{algorithmic}[1]
    \State Apply the variable-fixing on the parking spaces (call~\Cref{algo:subroutine} with~\(\matheuler{C}\) as input).
    \State Relax the integrality constraints of all binary variables in the~\ref*{formulation:third} model.
    \State Solve the relaxed model to obtain an optimal fractional solution.
    \LeftComment{The model is solved using the dual simplex method.} 
    \State Let~\(\scalebox{1.2}{$\bar{z}^{*}$}\) be the optimal value of the linear relaxation.
    \State Calculate the reduced costs~\(r_{c}\) associated with the variables~\(w_c\).
    \State Fix to zero the variables~\(\outgoing{x}^{\:\!p}_{\hspace*{-.2mm}d}, \outgoing{z}^{\:\!p}_{t}, \incoming{x}^{p}_{\hspace*{-.2mm}d}\) and~\(\incoming{z}^{p}_{t}\) if their value in the solution is zero.
    \LeftComment{This is done by setting the upper bound of these variables to zero.} 
    \State Restore the integrality constraints for the variables~\(\outgoing{x}^{\:\!p}_{\hspace*{-.2mm}d}\) and \(\incoming{x}^{p}_{\hspace*{-.2mm}d}\).
    \State Solve the resulting model for at most~\textsf{T}\(_{^\text{\textsf{MAXRUN}}}\) seconds.
    \If{not failed to find any integer feasible solution}
      \State Set \((\bar{\boldsymbol{w}}, \bar{\boldsymbol{x}}, \bar{\boldsymbol{y}}, \bar{\boldsymbol{z}}, \bar{\boldsymbol{\ell}})\) to be the best solution found during the solving process.
    \Else
      \State Build a feasible solution using the assignment plans construction heuristic of \cite{Silva/2023}.
      \State Set \((\bar{\boldsymbol{w}}, \bar{\boldsymbol{x}}, \bar{\boldsymbol{y}}, \bar{\boldsymbol{z}}, \bar{\boldsymbol{\ell}})\) to be the heuristic solution.
    \EndIf
    \State Let~\(\scalebox{1.2}{$\ubar{z}$}\) be the corresponding objective value of the solution \((\bar{\boldsymbol{w}}, \bar{\boldsymbol{x}}, \bar{\boldsymbol{y}}, \bar{\boldsymbol{z}}, \bar{\boldsymbol{\ell}})\).
    \State Apply the reduced cost variable-fixing as follows:%
            {\setlength{\abovedisplayskip}{1pt} \setlength{\abovedisplayshortskip}{1pt}%
             \setlength{\belowdisplayskip}{1pt} \setlength{\belowdisplayshortskip}{1pt}%
              \[w_c = 
                \begin{cases}
                  1\text{, if satisfies } r_{c} > \scalebox{1.2}{$\bar{z}$}^{*}\!-\scalebox{1.2}{$\ubar{z}$}\\
                  0\text{, if satisfies }\!-\!r_{c} > \scalebox{1.2}{$\bar{z}$}^{*}\!-\scalebox{1.2}{$\ubar{z}$}
                \end{cases}
              \]
            }
    \State Release the variables~\(\outgoing{x}^{\:\!p}_{\hspace*{-.2mm}d}, \outgoing{z}^{\:\!p}_{t}, \incoming{x}^{p}_{\hspace*{-.2mm}d}\) and~\(\incoming{z}^{p}_{t}\) previously fixed.
    \LeftComment{This is done by setting the upper bound of the fixed variables to one.} 
    \State Let~\mbox{\(\matheuler{C}^{\prime} \subseteq \matheuler{C}\)} denote the set of customers whose variables~\(w_{c}\) have not been fixed to zero.
    \State Apply the variable-fixing on the parking spaces (call~\Cref{algo:subroutine} with~\(\matheuler{C}^{\prime}\) as input).
    \State Repair the solution \((\bar{\boldsymbol{w}}, \bar{\boldsymbol{x}}, \bar{\boldsymbol{y}}, \bar{\boldsymbol{z}}, \bar{\boldsymbol{\ell}})\), if necessary.
    \State Solve the resulting model using \((\bar{\boldsymbol{w}}, \bar{\boldsymbol{x}}, \bar{\boldsymbol{y}}, \bar{\boldsymbol{z}}, \bar{\boldsymbol{\ell}})\) as a warm-start solution.
    \State Return the solution achieved.
  \end{algorithmic}
\end{algorithm}

Except for steps~\(1\), \(4\), and~\(5\), the~RCBVF algorithm's initial steps are the same as the~LRBVF heuristic.~But after step~\(7\), instead of optimally solving the resulting model, a maximum running time limit is imposed (\ie, the parameter~\textsf{T}\(_{^\text{\textsf{MAXRUN}}}\)) to focus the computational effort on the exploration of an optimal solution.~If the maximum running time limit is attained without a warm-start solution, then we apply (inside the `else' condition in step~\(12\)) the assignment plans construction heuristic described by~\cite{Silva/2023} to obtain an initial solution.~We are assuming that the warm-start solution obtained at step~\(8\) is better than the construction heuristic solution, and we have observed empirically that this assumption is true on the computational experiments.

After our two variable-fixing schemes are performed (steps~\(15\) and~\(18\)), the warm-start solution can be infeasible considering the set of variables associated with parking spaces in which their values were fixed to zero.~Then, the warm-start solution needs to be repaired (step~\(19\)).~This is easy when the current set of absent parking spaces with and without charging facilities, in the warm-start solution, can be replaced by empty parking spaces.~Nevertheless, some absent parking spaces without charging facilities cannot be replaced by others without charging facilities.~In this case, it should be noted that any optimal solution involving a vehicle parked (by a customer) at a parking space without a charging facility can be transformed into another feasible solution if there is an empty parking space with a charging facility available for at least the same~parking period.

Finally, the updated model is solved using a feasible warm-start solution (step~\(20\)), which is able to achieve an optimal solution. 

\begin{corollary}
  \label{corollary:RCBVF}
  The algorithm~{\normalfont\hyperref[algo:EVSP3-e]{RCBVF}} finds an optimal solution to any instance of the~{\normalfont\shortname}.
\end{corollary}

\section{Computational experiments} 
\label{sec:experiments}

In this section, we report and discuss the results of the computational experiments performed to test our proposed formulation and the derived approaches.~First, the~\RemoveSpaces{\ref{formulation:third}} results are compared with the results of the previously best-performing formulation in the literature.~After that, the results of~\RemoveSpaces{\ref*{formulation:third}}, \hyperref[algo:EVSP3-H]{LRBVF}, and~\hyperref[algo:EVSP3-e]{RCBVF} are compared among each other, and finally, further analysis is carried out on parameter settings that yield better performance.

Our computational experiments are carried out on an~\mbox{Ubuntu 18.04 64-bit} operating system with an~Intel(R) Xeon(R) E5-2630 v4 running at~\mbox{2.2 GHz}, \(10\)~cores, \mbox{and \(64\) GB RAM}.~The implementation was done in~{\CC} and uses~Gurobi Optimizer \mbox{version 9.0.3} as the mixed-integer linear programming solver, compiled with~GCC version 7 using the optimization flag~-O3.~We used Gurobi Optimizer with the default settings and restricted the computation to a single thread.~Unless explicitly stated otherwise, a time limit of one hour is set for solving the instances.



\subsection{Comparison with previous results}

In this first study, we conduct computational experiments on two benchmark sets proposed by~\cite{Silva/2023}.~One of the sets consists of random instances generated from a~\(50\)~km by~\(50\)~km grid, divided into ten groups of~\(30, 40, \dots,\) and~\(120\) customers, each with twenty instances, ranging from three to five stations (capacity of each station ranging from~\(10\) to~\(20\)), and each customer has at most four demands.~The other set consists of instances based on real-world data from an electric car-sharing system located in Fortaleza, Brazil, and it considers four scenarios, which are briefly summarized as follows:

\begin{enumerate}[label={\bf\Roman*}.]
  \item The first scenario, closest to the VAMO Fortaleza~\citep{Fortaleza/2018}, has two stations with a capacity of~\(3\) cars and ten stations with a capacity of~\(4\) cars, and all parking spaces are equipped with charging facilities.~Furthermore, each customer has only one demand.~In this scenario, the instances can be divided into ten groups of~\(250, 260, \dots,\) and~\(340\) customers.
  \item The second scenario was designed for the standard setup in which customers can have one or more driving demands (at most four).~In this scenario, the instances can be divided into ten groups of~\mbox{\(30, 40, \dots,\) and~\(120\)} customers.
  \item The third scenario increases the capacity of the stations by adjusting the number of parking spaces equipped and unequipped with charging facilities, as well as the number of vehicles, by modifying the generation parameters for the second scenario to be equal to those used in the grid-based instances.
  \item The last scenario is generated from the third one, in which the capacity of each station is increased to accommodate all vehicles in the car-sharing system. 
\end{enumerate}

Furthermore, we drop the integrality constraints for variables whose linear relaxation is guaranteed to lead to binary values (see~\Cref{remark:evsp3-relaxed} and~\Cref{proposition:EVSP3_integer}) —~these variables are treated as implicit binary variables, that is, variables that are required to be binary in any feasible solution but do not require explicit integrality constraints.~We also provide an initial solution as a warm-start for each execution, which is obtained using the construction heuristic proposed by~\cite{Silva/2023}.

We compare the proposed formulation to the previously best-performing one, namely~\RemoveSpaces{\makerefSb{formulation:second_splitted}}~(see \cite{Silva/2023}).~\Cref{tab:evsp3:summary} summarizes the computational results for both benchmark sets.~The percentage of optimal solutions obtained by their mixed-integer linear programs and the average time~(in seconds) to solve the instances are shown in this table.~We can see from these results that~\RemoveSpaces{\ref*{formulation:third}} outperforms~\RemoveSpaces{\ref*{formulation:second_splitted}}, giving an excellent performance for all the instances in a negligible computation time.

\begin{table}[H]
  \caption{Summary of overall results obtained from the two benchmark sets.}
  \label{tab:evsp3:summary}
  \setlength{\tabcolsep}{.1em}
  \renewcommand{\arraystretch}{1.1}
  \resizebox{\textwidth}{!}{%
  \begin{tabular}{rrrrrrrrrrrrrrrrrrr}
  \toprule
  \multicolumn{1}{c}{}       &                        & \multicolumn{2}{c}{\bf Grid-based}                       &                        & \multicolumn{11}{c}{\bf VAMO Fortaleza}                                                                                                                                                                                                                                                                              \\ \cline{3-4} \cline{6-16} 
                             &                        & \multirow{2}{*}{EVSP2-S}    & \multirow{2}{*}{~EVSP3}    &                        & \multicolumn{2}{c}{I}                                    &                        & \multicolumn{2}{c}{II}                                   &                        & \multicolumn{2}{c}{III}                                  &                        & \multicolumn{2}{c}{IV}                                   \\ \cline{6-7} \cline{9-10} \cline{12-13} \cline{15-16}
  \multicolumn{1}{c}{}       & \multicolumn{1}{c}{~~} &                             &                            & \multicolumn{1}{c}{~~} & \multicolumn{1}{c}{EVSP2-S} & \multicolumn{1}{c}{~EVSP3} & \multicolumn{1}{c}{~~} & \multicolumn{1}{c}{EVSP2-S} & \multicolumn{1}{c}{~EVSP3} & \multicolumn{1}{c}{~~} & \multicolumn{1}{c}{EVSP2-S} & \multicolumn{1}{c}{~EVSP3} & \multicolumn{1}{c}{~~} & \multicolumn{1}{c}{EVSP2-S} & \multicolumn{1}{c}{~EVSP3} \\ \hline
  \%Opt.                     &                        &     63.5\%                  &  100\%                     &                        &   100\%                     &   100\%                    &                        &    100\%                    &  100\%                     &                        &    15.5\%                   &  100\%                     &                        &    57.5\%                   & 100\%                      \\
  Avg. Time                  &                        & >1915.42\,s                 & 7.23\,s                    &                        & 45.08\,s                    & 0.46\,s                    &                        & 104.12\,s                   & 1.76\,s                    &                        & >3183.84\,s                 & 3.30\,s                    &                        & >1873.04\,s                 & 59.63\,s                   \\ \toprule
  \multicolumn{16}{l}{\small \emph{Note.} \%Opt., percentage of optimal solutions; Avg. Time, average of computational times (in seconds).}
  \end{tabular}}
\end{table}

To indicate the compactness of the formulation~\RemoveSpaces{\ref*{formulation:third}} compared to~\RemoveSpaces{\ref*{formulation:second_splitted}}, \Cref{fig:grid:dimension} shows the boxplot for the number of binary variables~(left plot) and the boxplot for the number of constraints~(right plot).~Each boxplot shows the median as a horizontal line inside the box, and the extremes of each box represent the first and third quartiles, respectively, while the whiskers represent the smallest and largest values.~In particular, to the number of binary variables, the blue boxplots are bigger than the red ones because Gurobi's presolve phase changes some relaxed variables to (implicit) binary variables since that continuous as well as binary variables can be declared as implicit binary variables.

\begin{figure}[H]
  \centering
  \begin{center}
  \begin{tikzpicture}[scale=0.83]
    \begin{axis}[box plot width=2mm, xtick distance=1, 
                  ylabel={Number of binary variables},
                  ylabel style={at={(0.08,0.5)}},
                  xmin=0,xmax=3,
                  xticklabels={,,
                                {\small \ref*{formulation:second_splitted}},
                                {\small \ref*{formulation:third}}
                              }
                  ]
      \boxplot [
        forget plot, gray, xshift=-.5cm,
        box plot whisker bottom index=1,
        box plot whisker top index=5,
        box plot box bottom index=2,
        box plot box top index=4,
        box plot median index=3
      ] {binary:original.dat}
      \boxplot [
        forget plot, red, xshift=.0cm,
        box plot whisker bottom index=1,
        box plot whisker top index=5,
        box plot box bottom index=2,
        box plot box top index=4,
        box plot median index=3
      ] {binary:relax.dat}
      \boxplot [
        forget plot, MidnightBlue, xshift=.5cm,
        box plot whisker bottom index=1,
        box plot whisker top index=5,
        box plot box bottom index=2,
        box plot box top index=4,
        box plot median index=3
      ] {binary:relax_and_presolved.dat}
    \end{axis}
  \end{tikzpicture}
  \qquad
  \begin{tikzpicture}[scale=0.83]
    \begin{axis}[box plot width=2mm, xtick distance=1, 
                  ylabel={Number of constraints},
                  ylabel style={at={(0.08,0.5)}},
                  xmin=0,xmax=3,
                  xticklabels={,,
                                {\small \ref*{formulation:second_splitted}},
                                {\small \ref*{formulation:third}}
                              }
                ]
      \boxplot [
        forget plot, red, xshift=-.25cm,
        box plot whisker bottom index=1,
        box plot whisker top index=5,
        box plot box bottom index=2,
        box plot box top index=4,
        box plot median index=3
      ] {constraints:relax.dat}
      \boxplot [
        forget plot, MidnightBlue, xshift=.25cm,
        box plot whisker bottom index=1,
        box plot whisker top index=5,
        box plot box bottom index=2,
        box plot box top index=4,
        box plot median index=3
      ] {constraints:relax_and_presolved.dat}
    \end{axis}
  \end{tikzpicture}
  \end{center}
  \caption{Distribution of the number of binary variables and constraints on the group with~\(120\) customers of the grid-based instances.~The red boxplots correspond to the mixed-integer linear programs obtained after integrality relaxation with respect to the implicit binary variables, while the blue boxplots correspond to the programs after presolve phase.~In addition, the gray boxplots correspond to the programs with any integrality relaxation.}
  \label{fig:grid:dimension}
\end{figure}
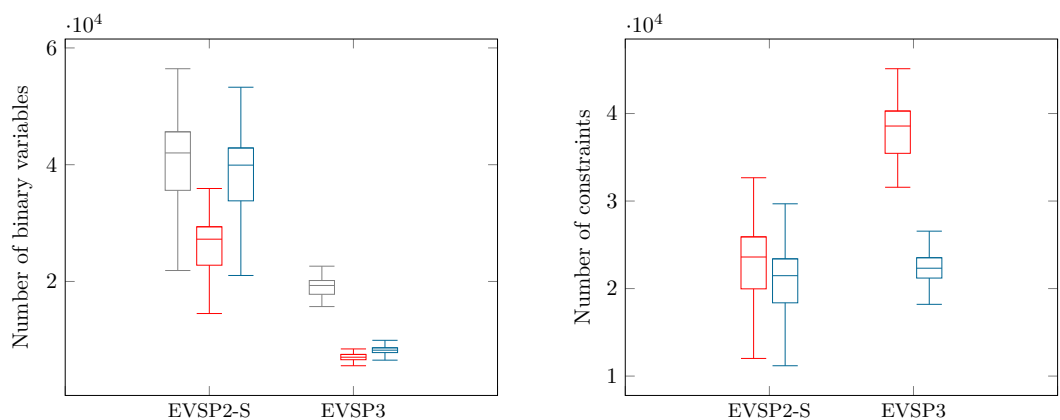

From those boxplots, we can see that~\RemoveSpaces{\ref*{formulation:third}} is more compact than~\RemoveSpaces{\ref*{formulation:second_splitted}} on the number of binary variables, but not on the number of constraints (on average, even after the presolve phase) —~it is worth mentioning that the same holds true for continuous variables.~Also, if the number of parking spaces is increased in a given instance, then the number of binary variables in~\RemoveSpaces{\ref*{formulation:third}} will be increased, but not in~\RemoveSpaces{\ref*{formulation:second_splitted}}.~Therefore, instances with a large number of parking spaces can be more challenging for formulation~\RemoveSpaces{\ref*{formulation:third}}.



\subsection{Results on new instances}

To compare the quality of the solutions found by~\RemoveSpaces{\ref{formulation:third}} to the ones found by~\hyperref[algo:EVSP3-H]{LRBVF} and~\hyperref[algo:EVSP3-e]{RCBVF} algorithms, we introduce new instances to the set VAMO Fortaleza, only for Scenarios~III and~IV, since they are more computationally expensive to compute.~Each of these scenarios has~\(200\) new instances, which can be divided into two subsets: one with groups of~\(280, 290, \dots,\) and~\(320\) customers, and another with groups of~\(480, 490, \dots,\) and~\(520\) customers (each group with~\(20\) instances).~All of these instances are available at~\href{https://gitlab.com/welverton/evsp}{https:{\small/\!/}gitlab.com{\small/}welverton{\small/}evsp}.~The setup for the formulation~\RemoveSpaces{\ref{formulation:third}} is the same as it was in the previous subsection, and the value of the~RCBVF algorithm's parameter~\textsf{T}\(_{^\text{\textsf{MAXRUN}}}\) is~\(12\) minutes.

\Cref{fig:vamo:evsp3_part_one} shows the percentage of the new instances that were solved to optimality.~Accordingly, out of the new~\(200\) instances for~Scenario~III, the proportion of solved instances was~\(63.5\%\) for~\RemoveSpaces{\ref{formulation:third}} and~\(62.5\%\) for~RCBVF.~It is also worth mentioning that both solved all the instances in the first subset optimally (\ie, with up to~\(320\) customers).~In the second plot, for~Scenario~IV, we can observe that the~RCBVF approach performs better than~\RemoveSpaces{\ref{formulation:third}}, and it seems to be preferable in cases with numerous parking spaces.~More specifically, the proportion of solved instances was~\(27.5\%\) and~\(48.5\%\), respectively.

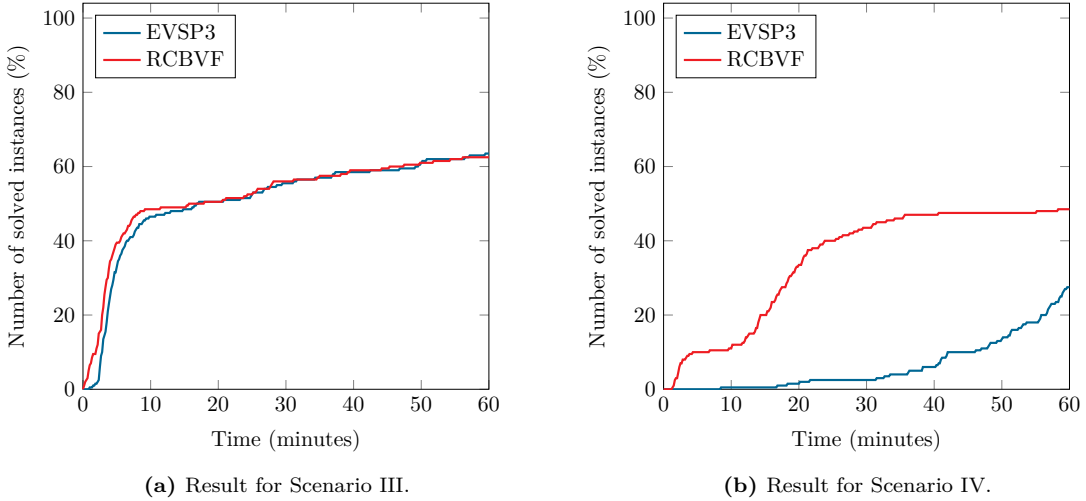
\begin{figure}[H]
  \centering
  \resizebox{\textwidth}{!}{%
  \captionsetup[subfigure]{oneside,margin={.5cm,0cm}}
  \begin{subfigure}[t]{.47\textwidth}
    \begin{tikzpicture}[scale=0.83]
      \begin{axis}[
            xlabel={Time (minutes)},
            ylabel={Number of solved instances (\%)},
            ylabel style={at={(0.03,0.5)}},
            legend entries={
              \RemoveSpaces{\ref*{formulation:third}},
              \RemoveSpaces{RCBVF},
            },
            legend pos=north west,
            legend cell align=left,
            x=0.1085cm, y=0.06cm,
            scale only axis, 
            ymin=0,ymax=104,xmin=0,xmax=60,
            enlargelimits=false,
        ]
        \addplot[mark=none, MidnightBlue, line width=1pt] table{data/VAMO/S-III/profile/EVSP3.dat};
        \addplot[mark=none, Red, line width=1pt] table{data/VAMO/S-III/profile/RCBVF.dat};
      \end{axis}
    \end{tikzpicture}
    \caption{Result for Scenario III.}
  \end{subfigure}
  \qquad
  \begin{subfigure}[t]{.47\textwidth}
    \begin{tikzpicture}[scale=0.83]
      \begin{axis}[
          xlabel={Time (minutes)},
          ylabel={Number of solved instances (\%)},
          ylabel style={at={(0.03,0.5)}},
          legend entries={
            \RemoveSpaces{\ref*{formulation:third}},
            \RemoveSpaces{RCBVF},
          },
          legend pos=north west,
          legend cell align=left,
          x=0.1085cm, y=0.06cm,
          scale only axis, 
          ymin=0,ymax=104,xmin=0,xmax=60,
          enlargelimits=false,
        ]
        \addplot[mark=none, MidnightBlue, line width=1pt] table{data/VAMO/S-IV/profile/EVSP3.dat};
        \addplot[mark=none, Red, line width=1pt] table{data/VAMO/S-IV/profile/RCBVF.dat};
      \end{axis}
    \end{tikzpicture}
    \caption{Result for Scenario IV.}
  \end{subfigure}
  }
  \caption{Percentage of the new instances solved optimally within the time limit of one hour.}
  \label{fig:vamo:evsp3_part_one}
\end{figure}

Concerning the quality of the achieved solutions for the instances not solved optimally in \mbox{Scenario III}, we observed that both obtained competitive results.~On average, the gap value is~\(0.34\%\) for~\RemoveSpaces{\ref{formulation:third}} and~\(0.37\%\) for~RCBVF.~However, in the case of~Scenario~IV, we observed that their average gap values are~\(167.95\%\) and~\(3.05\%\), respectively, meaning that (on average)~\RemoveSpaces{\ref{formulation:third}} does not perform well for instances not solved optimally.~Gap values were calculated as~\mbox{(\hspace*{-.1mm}(\hspace*{-.1mm}\textsf{UB} - \textsf{LB})/\textsf{LB})\(\times 100\)}, where \textsf{UB} and~\textsf{LB} are the upper and lower bounds achieved by the solver, respectively.

\Cref{fig:vamo:evsp3_part_two} shows the relative optimality gaps between the best upper bound value (\ie, minimum upper bound between~\RemoveSpaces{\ref{formulation:third}} and~RCBVF) and the solution value provided by the~LRBVF algorithm with a given time limit of~\(12\) minutes (dashed red lines) and~\(60\) minutes (blue lines).~These results show that the~LRBVF algorithm provides high-quality solutions within a short time, even when considering a large number of parking spaces.~It is worth mentioning that optimality gaps considering the solution value provided by assignment plans construction heuristic proposed by~\cite{Silva/2023} were at least~\(195\%\).~We conclude that the key challenge for the solver is in obtaining high-quality primal bounds, meaning that the~RCBVF performed a major role in closing the optimality gaps.

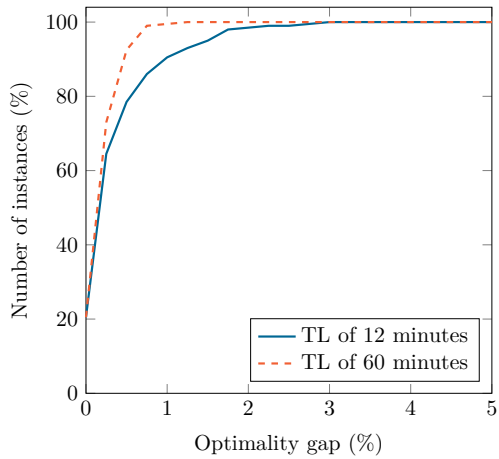
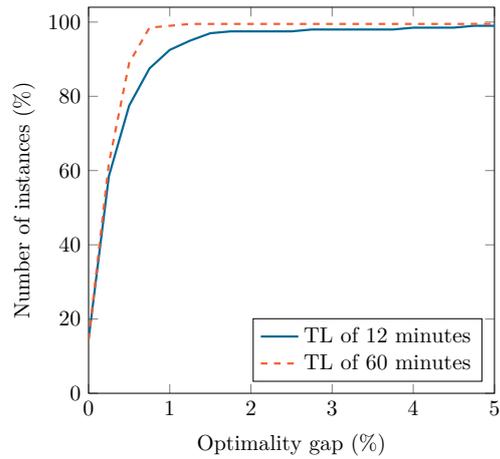
\begin{figure}[H]
  \centering
  \resizebox{\textwidth}{!}{%
  \captionsetup[subfigure]{oneside,margin={.5cm,0cm}}
  \begin{subfigure}[t]{.47\textwidth}
    \begin{tikzpicture}[scale=0.83]
      \begin{axis}[
            xlabel={Optimality gap (\%)},
            ylabel={Number of instances (\%)},
            ylabel style={at={(0.03,0.5)}},
            legend entries={
              \RemoveSpaces{TL of 12 minutes},
              \RemoveSpaces{TL of 60 minutes},
            },
            legend pos=south east,
            legend cell align=left,
            x=1.302cm, y=0.06cm,
            scale only axis, 
            ymin=0,ymax=104,xmin=0,xmax=5,
            enlargelimits=false,
        ]
        \addplot[mark=none, MidnightBlue, line width=1pt] table{data/VAMO/S-III/gap/EVSP3-H_12minutes.dat};
        \addplot[mark=none, dashed, RedOrange, line width=1pt] table{data/VAMO/S-III/gap/EVSP3-H_1hour.dat};
      \end{axis}
    \end{tikzpicture}
    \caption{Result for Scenario III.}
  \end{subfigure}
  \qquad
  \begin{subfigure}[t]{.47\textwidth}
    \begin{tikzpicture}[scale=0.83]
      \begin{axis}[
          xlabel={Optimality gap (\%)},
          ylabel={Number of instances (\%)},
          ylabel style={at={(0.03,0.5)}},
          legend entries={
            \RemoveSpaces{TL of 12 minutes},
            \RemoveSpaces{TL of 60 minutes},
          },
          legend pos=south east,
          legend cell align=left,
          x=1.302cm, y=0.06cm,
          scale only axis, 
          ymin=0,ymax=104,xmin=0,xmax=5,
          enlargelimits=false,
        ]
        \addplot[mark=none, MidnightBlue, line width=1pt] table{data/VAMO/S-IV/gap/EVSP3-H_12minutes.dat};
        \addplot[mark=none, dashed, RedOrange, line width=1pt] table{data/VAMO/S-IV/gap/EVSP3-H_1hour.dat};
      \end{axis}
    \end{tikzpicture}
    \caption{Result for Scenario IV.}
  \end{subfigure}
  }
  \caption{Percentage of relative optimality gaps between the best-known upper bound value and the solution value provided by the~LRBVF algorithm with a time limit (TL) of 12 and 60 minutes.~Gap values were calculated as~\mbox{(\hspace*{-.1mm}(\hspace*{-.1mm}\textsf{UB} - \textsf{LB})/\textsf{LB})\(\times 100\)}, where \textsf{UB} is the upper bound value and~\textsf{LB} is the solution value.}
  \label{fig:vamo:evsp3_part_two}
\end{figure}

\subsection{Further analysis}

We observed that the optimizer's default cuts are frequently too weak and do not produce better upper bounds.~The Gurobi optimizer allows access to parameters that control many algorithms for solving mixed-integer problems, such as~\textsf{Cuts} (which control the aggressiveness of cuts) and~\textsf{MIPFocus} (which control the computational effort between finding new solutions and proving optimality).~In this study, we investigate the impact of setting~\textsf{Cuts} to~\(0\) and~\textsf{MIPFocus} to~\(1\), which avoid the generation of any cuts and focus on finding new solutions, respectively.

\Cref{fig:vamo:evsp3_part_tree} shows the average ratio of the upper and lower bounds to the optimal solution value, using the default and new settings.~In both plots, the~\(y\)-axis value~\(1\) indicates that the objective bounds have reached the optimal value.~From these results, we can see that the solver using default settings takes a computational effort of almost~\(2\) times that of the solver using the proposed new settings. 

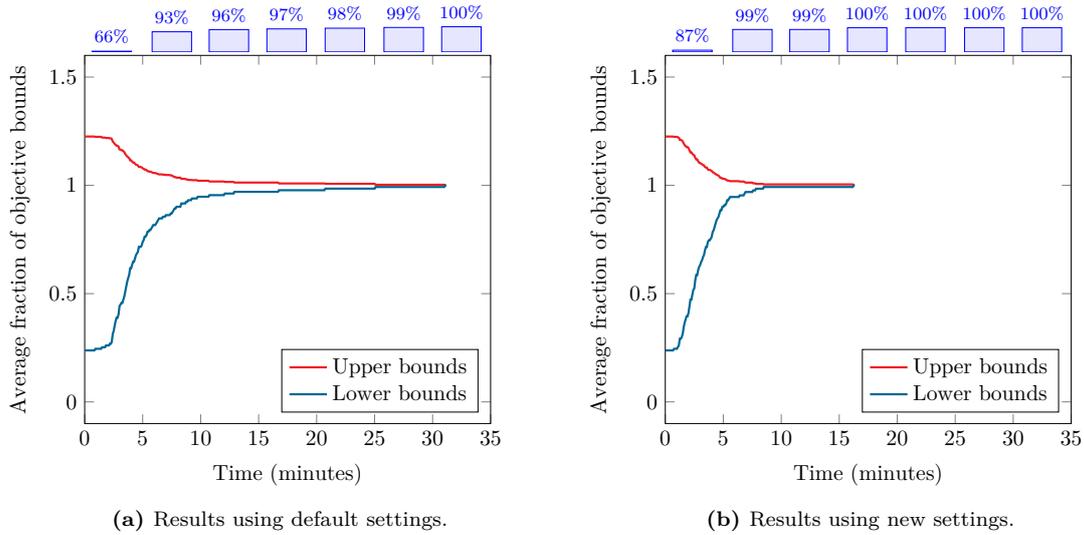
\begin{figure}[H]
  \centering
  \resizebox{\textwidth}{!}{%
  \captionsetup[subfigure]{oneside,margin={.5cm,0cm}}
  \begin{subfigure}[t]{.47\textwidth}
    \begin{tikzpicture}[scale=0.83]
      \begin{axis}[axis line style={opacity=0}, xtick=\empty, ytick=\empty,
                   nodes near coords={\pgfkeys{/pgf/fpu}\pgfmathparse{\pgfplotspointmeta*100}\footnotesize\pgfmathprintnumber{\pgfmathresult}\%},
                   enlarge x limits=false,
                   x=0.96348cm, xmin=0, xmax=8, ymax=1,
                   enlarge y limits={abs=0.01}, height=2cm
                  ]
        \addplot[ybar, bar width=18pt, blue, fill=blue!10] coordinates {
          (1+0.8, 0.66)
          (2+0.8, 0.93)
          (3+0.75, 0.96)
          (4+0.71, 0.97)
          (5+0.68, 0.98)
          (6+0.66, 0.99)
          (7+0.62, 1.00)
          };
      \end{axis}
      \end{tikzpicture}
      \begin{tikzpicture}[scale=0.83]  
      \begin{scope}[
          spy using outlines={
              rectangle,
              magnification=4,
              connect spies,
              size=2cm,
              blue,
          },
        ]
        \begin{axis}[
              xlabel={Time (minutes)},
              ylabel={Average fraction of objective bounds},
              ylabel style={at={(0.03,0.5)}},
              legend entries={
                Upper bounds,
                Lower bounds,
              },
              legend pos=south east,
              legend cell align=left,
              x=0.18599999999cm, y=3.5cm,
              scale only axis, 
              ymin=-0.1,ymax=1.6,xmin=0,xmax=35,
              enlargelimits=false,
          ]
          \addplot[mark=none, Red, line width=1pt] table{data/VAMO/S-III/conv/EVSP3_UB.dat};
          \addplot[mark=none, MidnightBlue, line width=1pt] table{data/VAMO/S-III/conv/EVSP3_LB.dat};
        \end{axis}
      \end{scope}
    \end{tikzpicture}
    \caption{Results using default settings.}
  \end{subfigure}
  \qquad
  \begin{subfigure}[t]{.47\textwidth}
    \begin{tikzpicture}[scale=0.83]
      \begin{axis}[axis line style={opacity=0}, xtick=\empty, ytick=\empty,
              nodes near coords={\pgfkeys{/pgf/fpu}\pgfmathparse{\pgfplotspointmeta*100}\footnotesize\pgfmathprintnumber{\pgfmathresult}\%},
              enlarge x limits=false,
              x=0.96348cm, xmin=0, xmax=8, ymax=1,
              enlarge y limits={abs=0.01}, height=2cm
            ]
      \addplot[ybar, bar width=18pt, blue, fill=blue!10] coordinates {
        (1+0.8, 0.87)
        (2+0.8, 0.99)
        (3+0.75, 0.99)
        (4+0.71, 1.00)
        (5+0.68, 1.00)
        (6+0.66, 1.00)
        (7+0.62, 1.00)
        };
      \end{axis}
    \end{tikzpicture}
    \begin{tikzpicture}[scale=0.83]
      \begin{scope}[
          spy using outlines={
              rectangle,
              magnification=4,
              connect spies,
              size=2cm,
              blue,
          },
        ]
        \begin{axis}[
            xlabel={Time (minutes)},
            ylabel={Average fraction of objective bounds},
            ylabel style={at={(0.03,0.5)}},
            legend entries={
              Upper bounds,
              Lower bounds,
            },
            legend pos=south east,
            legend cell align=left,
            x=0.18599999999cm, y=3.5cm,
            scale only axis, 
            ymin=-0.1,ymax=1.6,xmin=0,xmax=35,
            enlargelimits=false,
            xtick distance=5,
          ]
          \addplot[mark=none, Red, line width=1pt] table{data/VAMO/S-III/conv/EVSP3_UB+SpeedUp.dat};
          \addplot[mark=none, MidnightBlue, line width=1pt] table{data/VAMO/S-III/conv/EVSP3_LB+SpeedUp.dat};
        \end{axis}
      \end{scope}
    \end{tikzpicture}
    \caption{Results using new settings.}
  \end{subfigure}
  }
  \caption{The average ratio of objective bounds to the optimal value on the first subset~(\ie, with up to~\(320\) customers) of~Scenario~III.~In addition, each bar chart displays the percentage (as a cumulative histogram) of instances that were solved optimally.}
  \label{fig:vamo:evsp3_part_tree}
\end{figure}

\Cref{fig:vamo:evsp3_part_four} shows the percentage of the new instances that were optimally solved within one hour.~With the new settings, we can observe on the chart for~Scenario~III a slight performance boost.~However, for~Scenario~IV, a very significant performance gain can be observed.~From the results we can observe that the proportion of solved instances for the default setting was~\(27.5\%\) for~\RemoveSpaces{\ref{formulation:third}} and~\(48.5\%\) for the~RCBVF algorithm, whereas for the new settings their percentage of solved instances was~\(45.0\%\) and~\(49.5\%\), respectively.~It is worth mentioning that their average gap values for the instances that were not solved optimally are~\(233.37\%\) and~\(2.91\%\), respectively.

\begin{figure}[H]
  \centering
  \resizebox{\textwidth}{!}{%
  \captionsetup[subfigure]{oneside,margin={.5cm,0cm}}
  \begin{subfigure}[t]{.47\textwidth}
    \begin{tikzpicture}[scale=0.83]
      \begin{axis}[
            xlabel={Time (minutes)},
            ylabel={Number of solved instances (\%)},
            ylabel style={at={(0.03,0.5)}},
            legend entries={
              ,\RemoveSpaces{\ref*{formulation:third}},
              ,\RemoveSpaces{RCBVF},
            },
            legend pos=north west,
            legend cell align=left,
            x=0.1085cm, y=0.06cm,
            scale only axis, 
            ymin=0,ymax=104,xmin=0,xmax=60,
            enlargelimits=false,
            legend image code/.code={%
              \draw[dotted,opacity=0.5] (0cm,-0.08cm) -- (0.57cm,-0.08cm);
              \draw[solid] (0cm, 0.08cm) -- (0.57cm, 0.08cm);
            },
        ]
        \addplot[mark=none, dotted, MidnightBlue!50, line width=1pt] table{data/VAMO/S-III/profile/EVSP3.dat};
        \addplot[mark=none, MidnightBlue, line width=1pt] table{data/VAMO/S-III/profile/EVSP3+SpeedUp.dat};
        \addplot[mark=none, dotted, Red!50, line width=1pt] table{data/VAMO/S-III/profile/RCBVF.dat};
        \addplot[mark=none, Red, line width=1pt] table{data/VAMO/S-III/profile/RCBVF+SpeedUp.dat};
      \end{axis}
    \end{tikzpicture}
    \caption{Result for Scenario III.}
  \end{subfigure}
  \qquad
  \begin{subfigure}[t]{.47\textwidth}
    \begin{tikzpicture}[scale=0.83]
      \begin{axis}[
          xlabel={Time (minutes)},
          ylabel={Number of solved instances (\%)},
          ylabel style={at={(0.03,0.5)}},
          legend entries={
           ,\RemoveSpaces{\ref*{formulation:third}},
           ,\RemoveSpaces{RCBVF},
          },
          legend pos=north west,
          legend cell align=left,
          x=0.1085cm, y=0.06cm,
          scale only axis, 
          ymin=0,ymax=104,xmin=0,xmax=60,
          enlargelimits=false,
          legend image code/.code={%
            \draw[dotted,opacity=0.5] (0cm,-0.08cm) -- (0.57cm,-0.08cm);
            \draw[solid] (0cm, 0.08cm) -- (0.57cm, 0.08cm);
          },
        ]
        \addplot[mark=none, dotted, MidnightBlue!40, line width=1pt] table{data/VAMO/S-IV/profile/EVSP3.dat};
        \addplot[mark=none, MidnightBlue, line width=1pt] table{data/VAMO/S-IV/profile/EVSP3+SpeedUp.dat};
        \addplot[mark=none, dotted, Red!40, line width=1pt] table{data/VAMO/S-IV/profile/RCBVF.dat};
        \addplot[mark=none, Red, line width=1pt] table{data/VAMO/S-IV/profile/RCBVF+SpeedUp.dat};
      \end{axis}
    \end{tikzpicture}
    \caption{Result for Scenario IV.}
  \end{subfigure}
  }
  \caption{Percentage of the new instances solved optimally within the time limit of one hour using new settings (solid lines).~In particular, dotted lines indicate the previous results obtained using default settings.}
  \label{fig:vamo:evsp3_part_four}
\end{figure}
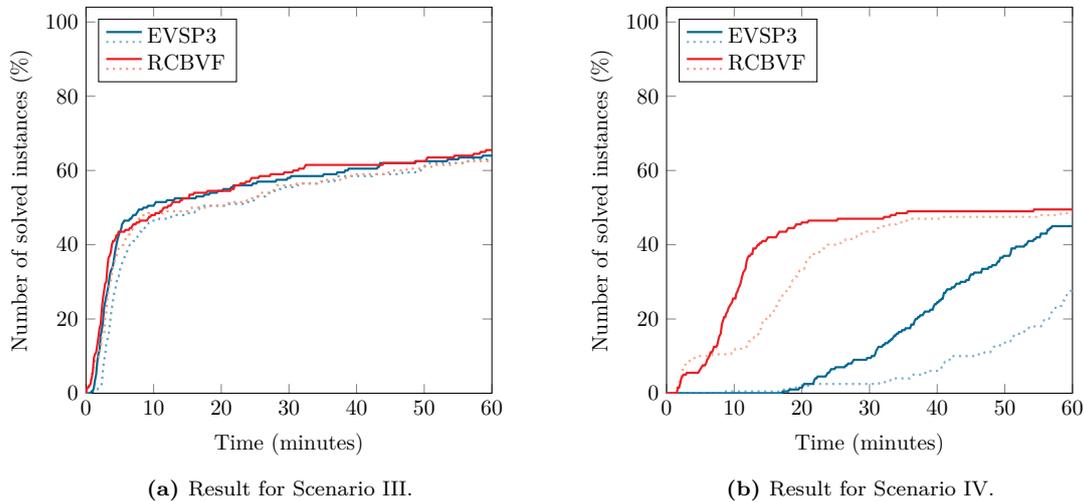

\section{Conclusion}
\label{sec:conclusions}

In this paper, we showed that, assuming~\mbox{P \(\neq\) NP}, for any~\mbox{\(\epsilon > 0\)}, the electric vehicle sharing problem~(\shortname) cannot be approximated in polynomial time within a factor of~\(n^{1-\epsilon}\)\!, where~\(n\) is the number of customers.~We also showed that the problem does not have a monotone structure, meaning it could impair constructive heuristics that rely on maintaining feasibility during a bottom-up (or top-down) search.~We presented a new mixed-integer linear programming formulation for the~\shortname, based on an energy flow modeling, referred to as~\RemoveSpaces{\ref{formulation:third}}.~A key feature to note in this modeling is that the vehicles are abstracted in the assignment of demands (\ie, not indexed by them).~Instead of binary flows in space-time networks, we model the energy of the vehicles as continuous flows.~In addition, we also presented two methods for solving instances of the~\shortname, both derived from the formulation~\RemoveSpaces{\ref*{formulation:third}}.~Our first method, referred to as~\hyperref[algo:EVSP3-H]{LRBVF}, is a heuristic algorithm based on fixing the zero-valued variables in a relaxed solution, which requires little coding effort, reducing the model size, to find a high-quality solution.~Our second method, referred to as~~\hyperref[algo:EVSP3-e]{RCBVF}, uses a reduced-cost variable-fixing strategy and another variable-fixing associated with parking spaces to find an optimal solution with a reduced computational effort.

The effectiveness of our approaches was thoroughly investigated by computational experiments.~First, we performed tests on two benchmark instances, and the outcomes were compared with the previous best-performing formulation~\citep{Silva/2023}, namely~\RemoveSpaces{\ref{formulation:second_splitted}}.~According to the results, the formulation~\RemoveSpaces{\ref*{formulation:third}} outperforms~\RemoveSpaces{\ref*{formulation:second_splitted}} in terms of optimal solutions and computation time.~Unsolved instances by~\RemoveSpaces{\ref*{formulation:second_splitted}} within a time limit of one hour, which is~\(32.7\%\) of all the instances used in the first study, were solved by~\RemoveSpaces{\ref*{formulation:third}} in less than~\(6\) minutes.~It is worth mentioning that, on average, the~computational effort of~\RemoveSpaces{\ref*{formulation:third}} was almost~\(15\) seconds.

Two new sets of instances were created to include more challenging scenarios to compare the quality of the solutions found by the formulation~EVSP3 to those found by~LRBVF and~RCBVF algorithms.~The results obtained for algorithms~LRBVF and~RCBVF show that they are computationally efficient and that they can achieve high-quality solutions in a short time, even for the most difficult scenario.~As shown by the results for new instances in the scenario where each station has the capacity to accommodate all vehicles in the car-sharing system, the~RCBVF algorithm was able to solve~\(1.72\) times more instances than the~formulation~\RemoveSpaces{\ref*{formulation:third}}.~Additionally, a third computational experiment was also carried out without using the solver's default settings, which nearly cut the computation time in half.

Future study could focus on matheuristic methods which might be used to solve large-scale real-world instances.~More specifically, future research could concentrate on decomposition approaches, which are based on the idea of identifying smaller, easier-to-solve subproblems than the original problem.




\bibliographystyle{abbrvnat}
\bibliography{references}

\end{document}